\documentclass[reprint, 
]{revtex4-2}
\usepackage[utf8]{inputenc}
\usepackage[english]{babel}

\usepackage[plainpages = false, pdfpagelabels, bookmarks, bookmarksopen = true, bookmarksnumbered = true, breaklinks = true, linktocpage, colorlinks = true, linkcolor = blue, urlcolor  = blue, citecolor = blue,  anchorcolor = green, hyperindex = true,  hyperfigures]{hyperref}

\usepackage{enumerate}
\usepackage{tabularx}
\usepackage{makecell, multirow}
\usepackage{tikz}
\usetikzlibrary{patterns}
\usepackage{slashed}
\usepackage{physics}
\usepackage{verbatim}
\usepackage{svg}
\usepackage{mathtools}
\usepackage{amssymb}
\usepackage{cancel}
\usepackage{bbold}
\usepackage{bm}
\usepackage{booktabs}
\usepackage[caption=false,singlelinecheck=off]{subfig}
\usepackage{blindtext}
\usepackage{bbm}
\usepackage{graphicx}
\usepackage{dcolumn}
\usepackage{bm}
\usepackage{orcidlink}
\usepackage{natbib}
\usepackage[normalem]{ulem}
\usepackage{xcolor}

\begin{document}
\title{\texorpdfstring{Nonuniform superconducting states from Majorana flat bands}{}}

\author{{Sushanth Varada}\,\orcidlink{0000-0002-7797-0385}}
\email{sushanth.varada@physics.uu.se}
\affiliation{Department of Physics and Astronomy, Uppsala University, Box 524, S-751 20 Uppsala, Sweden}

\author{{Aksel Kobia\l{}ka}\,\orcidlink{0000-0003-2881-8253}}
\affiliation{Department of Physics and Astronomy, Uppsala University, Box 524, S-751 20 Uppsala, Sweden}

\author{{Ankita Bhattacharya}\,\orcidlink{0000-0002-7432-9533}}
\affiliation{Department of Physics and Astronomy, Uppsala University, Box 524, S-751 20 Uppsala, Sweden}

\author{{Patric Holmvall}\,\orcidlink{0000-0002-1866-2788}}
\affiliation{Department of Physics and Astronomy, Uppsala University, Box 524, S-751 20 Uppsala, Sweden}

\author{{Annica M. Black-Schaffer}\,\orcidlink{0000-0002-4726-5247}}
\affiliation{Department of Physics and Astronomy, Uppsala University, Box 524, S-751 20 Uppsala, Sweden}

\date{\today}
\begin{abstract}
Zero-energy flat bands within the superconducting gap can give rise to competing ordered phases. We investigate such phases in topological superconductors based on the magnetic adatom platform hosting a flat band of Majorana edge states. 
Our self-consistent calculations of the superconducting order parameter show the emergence of both a pair density wave with edge-localized amplitude modulations, often a bond wave order, and a phase crystal characterized by longer-wavelength edge-localized phase modulations. 
Both these states lower the free energy by gapping out the Majorana flat band, but use different mechanisms tied to the underlying topology. 
In fact, at zero temperature the uniform superconducting solution with Majorana flat band never survives. Instead, a pair density wave appears when sufficient Majorana modes exist with oppositely signed topological winding numbers, while a phase crystal appears in regions of the phase diagram where one non-trivial winding number is heavily dominating. 
A crossover region connects these two states with modulations in both amplitude and phase. 
At finite temperatures, the pair density wave survives up to around 80\% of the bulk superconducting transition temperature, while the phase crystal only appears at lower temperatures and the crossover region is also strongly suppressed. 
Our findings establish the ubiquity of emergent nonuniform superconducting phases and their temperature-dependent behavior in topological superconductors.
\end{abstract}
\maketitle

\section{Introduction}\label{sec:Introduction}
Nonuniform superconducting ground states provide a fascinating platform to study the interplay between spontaneous symmetry breaking and ordered states~\cite{Abrikosov:1956sx,PhysRev.135.A550,Larkin:1964wok,PhysRevLett.88.117001, PhysRevLett.99.127003,Agterberg2008,PhysRevB.79.064515,PhysRevLett.122.165302}.
In particular, flat bands of zero-energy states within a superconductor can easily become thermodynamically unstable to interactions due to their extensive ground-state degeneracy, facilitating a competition between different symmetry-breaking instabilities~\cite{Vorontsov2018}. This has been extensively studied at the $(110)$ surface of $d$-wave superconductors, especially in the context of cuprate high-temperature superconductors~\cite{Vorontsov2018,Hu1994,Matsumoto1996,PhysRevLett.79.281,Sigrist1998,PhysRevB.96.060503, PhysRevB.101.235103, PhysRevB.62.R14653,Honerkamp2000, potter_edge_2014,PhysRevLett.110.197001, Holmvall2018,PhysRevB.99.184511,Hakansson2015,PhysRevResearch.2.013104, PhysRevResearch.2.043198, PhysRevB.110.064502,5bl6-g7x4,sato2026revisitingsurfacedensitystates}. Notably, different superconducting states can occur that lower the free energy by shifting the zero-energy flat bands to finite energies, including a subdominant superconducting order~\cite{Matsumoto1996,Sigrist1998,PhysRevLett.79.281,PhysRevB.53.2835,Tanuma1998,PhysRevB.96.060503,PhysRevB.101.235103}, spontaneous but translationally invariant supercurrents~\cite{Higashitani1997Apr,PhysRevLett.82.3336,Barash2000Sep,PhysRevB.62.R14653,Vorontsov2009Apr,PhysRevB.89.184508}, and edge ferromagnetism~\cite{Honerkamp2000,potter_edge_2014}. However, the most energetically favorable scenario appears to be the superconducting phase crystal~\cite{Hakansson2015,PhysRevResearch.2.013104}, a nonuniform superconducting ground state characterized by periodic spontaneous phase gradients. Furthermore, the phase crystal state has gained attention because it remains robust in the presence of strong correlations, disorder, finite-temperature effects~\cite{Chakraborty2022, PhysRevB.109.L180509, PhysRevB.111.094513} and provides good agreement with cuprate experiments~\cite{Chakraborty2022,PhysRevLett.79.277,PhysRevB.55.R14757, PhysRevLett.81.2542, PhysRevLett.87.177004, PhysRevLett.88.047001}. 

In this work, we investigate prospects for the emergence of nonuniform superconducting ground states, including the phase crystal, in a topological superconductor hosting flat bands of topologically protected Majorana states. This addresses the interplay between, on one hand, forming non-uniform superconducting states to lower the energy, and, on the other hand, the topological protection of the Majorana states. 
In particular, we are interested in understanding how the topological protection affects, or possibly hinders, the emergence of nonuniform superconducting phases. 
Similar avenues have previously been explored in the context of stability of Majorana states in stacked nanowires~\cite{Li_2013,PhysRevB.89.174514} and small-scale cold-atom systems~\cite{MizushimaSato2013,wang_fate_2014,PhysRevA.92.023621, PhysRevLett.121.185302}. So-called edge-induced pairing modulations in amplitude~\cite{PhysRevA.92.023621} and phase~\cite{MizushimaSato2013, wang_fate_2014} of the superconducting order parameter have been shown to break symmetries and hybridize Majorana states. Here, we identify these modulations as two different nonuniform superconducting phases, namely, a pair density wave~\cite{PhysRevB.79.064515,Agterberg2008,pdw_footnote} and a phase crystal~\cite{PhysRevResearch.2.013104}, and characterize these two states through energy-based arguments and their connection to topology.  

\begin{figure} [!t]
    \centering
    \includegraphics[width=0.9\linewidth]{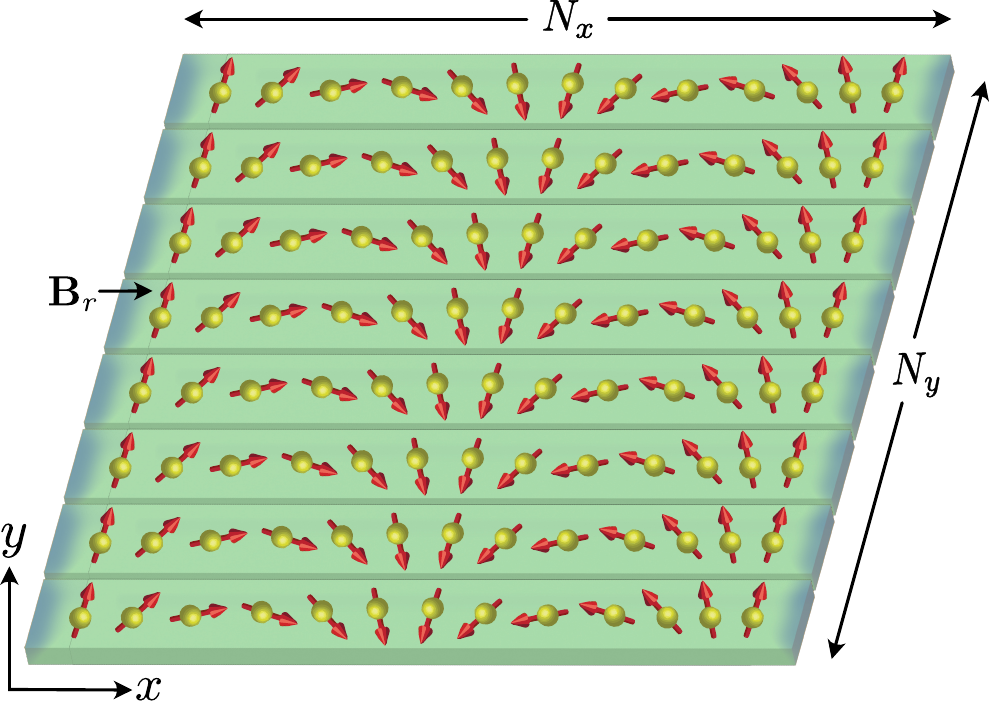}
    \caption{Schematic illustration of the experimental setup to obtain \mbox{Majorana} flat bands. Magnetic atoms (yellow spheres) with field strength $\mathbf{B}_r$ (red arrows) on an $s$-wave superconducting substrate (green) are tuned to form coplanar spin spirals with the pitch $\mathbf{q}=(q_x,0)$. For a wide range of parameters the system enters a topologically non-trivial regime hosting multiple Majorana edge states along the $y$-direction (schematically illustrated with gray colored cloud). The system consists of $N_x \times N_y$ sites.}
    \label{fig:2D_lattice_schematic}
\end{figure}

To this end, we consider a Majorana flat band model based on atomic spin chains with a helical magnetic texture deposited on an $s$-wave superconducting substrate~\cite{PhysRevB.84.195442,PhysRevB.85.144505,PhysRevB.88.155420,PhysRevB.90.060401,PhysRevB.93.140503,PhysRevB.88.020407,PhysRevB.89.115109, sedlmayr_flat_2015}, see Fig.~\ref{fig:2D_lattice_schematic}. Arrays of such magnetic adatoms have been shown to lead to multiple spin polarized sub-gap states that hybridize to form bands~\cite{Balatsky_2006}. These bands can enter a topologically non-trivial superconducting phase~\cite{PhysRevLett.114.236803,Li2016,Schneider2021, realistic_2D_Kitaev_model} supporting flat bands of Majorana edge states protected by chiral symmetry~\cite{sedlmayr_flat_2015, sedlmayr2015majoranas}. Developments in single-atom manipulation techniques based on scanning tunneling microscopy (STM) have enabled the fabrication, detection, and control of these superconducting sub-gap states~\cite{Heinrich2018, PhysRevLett.100.226801, PhysRevLett.115.197204,Kamlapure2018,Yang2019,PhysRevB.104.045406}. This offers a promising experimental route to realize topological superconductivity in atomic spin chains on a superconductor~\cite{rachel_majorana_2025,Nadj-Perge2014,Kim2018,Schneider2020,Schneider2022,Crawford2022,Soldini2023,Bruningetal2025}. Specifically, we focus on systems with non-collinear spin spirals in proximity to a conventional superconducting substrate~\cite{Kim2018,Schneider2020,Bruningetal2025}. To identify the possible ground states in such systems, we start from a tight-binding model for coplanar spin spirals on a 2D $s$-wave superconductor~\cite{sedlmayr_flat_2015} and perform self-consistent calculations of the effective superconducting order parameter to find the ground state. 

In more detail, we find that the Majorana flat band edge states, present for a uniform superconducting state, are always unstable at low temperatures towards the formation of either a pair density wave or a phase crystal along the edges of the system, or the combination thereof. The pair density wave exhibits edge localized superconducting amplitude modulations without any spatial variation in phase. The pair density wave predominantly displays short-range modulations on the atomic scale, even approaching a bond wave order~\cite{wang_fate_2014}. In contrast, the phase crystal is characterized by strong superconducting phase modulations at the edges on the superconducting coherence length scale, which drive spontaneous current loops and with only weak modulations in amplitude set by the phase modulation length scale. Both of these nonuniform superconducting states split and shift the Majorana flat band away from zero energy, thereby lowering the overall free energy of the system, but using different mechanisms. 
The pair density wave appears when sufficient Majorana modes exist with oppositely signed winding numbers, which allow for amplitude modulations through band folding, while the phase crystal appears when one non-trivial winding number is dominating and thus instead phase modulations are needed to gap out the Majorana flat band.
Thus, topology, set primarily by the chemical potential, determines which nonuniform state is formed.
This results in an experimentally accessible tunability between the pair density wave and the phase crystal. 

We find that the pair density wave and phase crystal are roughly as abundant as ground states in the phase diagram, where the order parameter only develops phase modulations when amplitude modulations are not sufficient to shift the Majorana states away from zero energy. In the zero-temperature limit, these two phases are connected by a crossover, where the superconducting order parameter exhibits comparable modulations in amplitude and phase. However, this crossover region diminishes at finite temperatures. The pair density wave also typically exhibits a higher transition temperature, reaching up to around 80\% of the mean-field bulk superconducting transition temperature $T_c$, while the phase crystal survives up to around 10\% of $T_c$. We trace this behavior to the phase crystal only emerging in a parameter regime where both the number of Majorana states and the magnitude of the bulk superconducting gap are relatively small, making it more sensitive to thermal excitations. Above the transition temperatures of these nonuniform superconducting states, the uniform superconducting state with its Majorana flat bands reappears.

The remainder of this work is organized as follows. We discuss the model of the Majorana flat band system and methods used in Sec.~\ref{sec:Model}. In Sec.~\ref{sec:Phase_Diagrams}, we characterize the nonuniform superconducting phases emerging in the zero-temperature limit. This is followed by an analysis of the thermal stability of these phases in Sec.~\ref{sec:Finite_temperature_phase_diagram}. We summarize our results in Sec.~\ref{sec:Summary_Conclusions}. Appendices~\ref{app:gauge_transformation} and ~\ref{app:winding_number} contain details of our derivations, while Appendix~\ref{app:additional_bwo_pc_data} provides additional numerical data.   
\enlargethispage{1\baselineskip}
\section{Model and methods}\label{sec:Model}
\subsection{Magnetic adatoms on conventional superconductor}
We consider an array of magnetic impurities deposited onto a conventional $s$-wave superconducting substrate, as illustrated in Fig.~\ref{fig:2D_lattice_schematic}. The magnetic adatoms form a spin polarized band within the superconducting gap, which can be tuned into a topological phase hosting Majorana flat bands~\cite{PhysRevB.88.020407,PhysRevB.88.155420,PhysRevB.89.115109,sedlmayr_flat_2015,Bruningetal2025}. The two-dimensional (2D) tight-binding Bogoliubov-de Gennes (BdG) model Hamiltonian of this system in the Nambu basis \mbox{$\Psi = (c_{r\uparrow}, c_{r\downarrow}, c^{\dagger}_{r\downarrow}, -c^{\dagger}_{r\uparrow})^T$} is given by \cite{sedlmayr_flat_2015}
\begin{align}\label{eq:MFB_stability_true_H}
    \Tilde{H} &= \sum_{r} \Psi_r^{\dagger}[-\mu\,\tau_z\sigma_0+\tau_0\,\mathbf{B}_r\cdot\boldsymbol{\sigma}]\Psi_r \nonumber \\
    & \hspace{0.5cm} +\sum_{r} \Psi_r^{\dagger}\left[-\Delta_r \,\frac{1}{2}(\tau_x+i\tau_y)\sigma_0 + \text{H.c.}\right]\Psi_r \nonumber\\
    & \hspace{0.8cm} +\sum_{\left<r,r'\right>}\Psi^{\dagger}_r\left[-\frac{t}{2}\,\tau_z\sigma_0\right]\Psi_{r'} \,,
\end{align}
where $c^{\dagger}_{r\sigma}$ creates an electron on the $r$-th lattice site with spin $\sigma=\uparrow,\downarrow$. Here, $\boldsymbol{\tau}$ and $\boldsymbol{\sigma}$ denote the Pauli matrices acting in particle-hole and spin subspaces, respectively. Furthermore, $\mu$ is the chemical potential and \mbox{$\mathbf{B}_r = B(\cos\phi_r\sin\varphi_r, \sin\phi_r\sin\varphi_r,\cos\varphi_r)$} represents the effective inhomogeneous magnetic field produced by the magnetic impurities. Here, $B$ is the effective field strength, while $\phi_r$ and $\varphi_r$ denote the site-dependent in-plane and out-of-plane angles, respectively. The hopping between the nearest neighbor lattice sites $\langle r,r'\rangle$ is characterized by $t$. The $s-$wave superconducting order $\Delta_r$, induced by the substrate is a complex quantity with amplitude $|\Delta_r|$ and phase $\theta_r$. The superconducting order parameter is determined self-consistently \cite{zhu_bogoliubov-gennes_2016} via
\begin{align}\label{eq:self_consistent_condition}
    \Delta_r &= \frac{V}{2}\big[\langle c_{r\uparrow} c_{r\downarrow}\rangle - \langle c_{r\downarrow} c_{r\uparrow}\rangle\big],\nonumber\\
    & = \frac{V}{2}\sum_{E_{n}>0} \left[u^n_{r\uparrow}v^{n*}_{r\downarrow} + u^n_{r\downarrow}v^{n*}_{r\uparrow}\right]\tanh\left(\frac{E_n}{2k_BT}\right). 
\end{align}
Here, $V>0$ denotes a constant attractive interaction, needed to generate the conventional spin-singlet $s$-wave superconducting state present in the absence of magnetic impurities,  while $E_n$ is the eigenenergy corresponding to the $n$-th eigenstate $(u^n_{r\uparrow}, u^n_{r\downarrow}, v^n_{r\downarrow}, -v^n_{r\uparrow})^T$ of the BdG Hamiltonian in Eq.~\eqref{eq:MFB_stability_true_H}. Furthermore, $T$ is the temperature and $k_B$ is the Boltzmann constant. 

We consider a general class of spin textures on the superconducting substrate with magnetic moments forming coplanar non-collinear spin spirals, as illustrated in Fig.~\ref{fig:2D_lattice_schematic} and inspired by recent experimental progress \cite{Kim2018,Schneider2020,Bruningetal2025}. This setup generates an in-plane magnetic field with $\varphi_r=\pi/2$, and \mbox{$\phi_r=2\pi\,\mathbf{q}\cdot\mathbf{r}$}, where $\mathbf{q}=(q_x,q_y)$ denotes the spiral pitch vector. Experimentally, the spiral pitch is a site-dependent quantity, prone to inhomogeneities due to magnetic anisotropy~\cite{Bruningetal2025}. While mild deviations would retain the non-collinear characteristics of the spin spiral, severe inhomogeneity may be detrimental to the formation of a Majorana flat band. Here, since we are interested in effects due to the existence of a Majorana flat band, we assume a homogeneous spiral pitch as a tunable parameter independent of lattice sites~\cite{sedlmayr_flat_2015}. 

Next, we perform a local gauge transformation~\cite{PhysRevB.85.144505,realistic_2D_Kitaev_model,zhu_bogoliubov-gennes_2016} to remove the site-dependence of the magnetic field (see  Appendix~\ref{app:gauge_transformation} for details) resulting in  
\begin{align}\label{eq:gauge_transformed_H}
    H &= \sum_{r} \Psi_r^{\dagger}\left[-\mu\,\tau_z\sigma_0+B\,\tau_0\sigma_x\right]\Psi_r \nonumber\\ 
    &\hspace{0.5cm}+\sum_{r} \Psi_r^{\dagger}\left[-\Delta_r \,\frac{1}{2}(\tau_x+i\tau_y)\sigma_0 + \text{H.c.}\right]\Psi_r \nonumber\\ 
    &\hspace{0.8cm} + \sum_{\left<r,r'\right>}\Psi^{\dagger}_r\left[-\frac{t}{2}\,\tau_z\,\exp(i\sigma_z\pi\,\mathbf{q}\cdot\mathbf{v}_{r,r'})\right]\Psi_{r'}\,.
\end{align}
Here, $\mathbf{v}_{r,r'} = (\delta_{r_{x}+1,r'_x},\delta_{r_{y}+1,r'_y})$ is the real-space vector between the nearest neighbor sites. The exponential phase factor in the kinetic energy generates an effective spin-orbit interaction by coupling the orbital motion to the spin. Importantly, its strength is governed by the spiral pitch $\mathbf{q}$, and unlike the conventional relativistic spin-orbit interaction, this effective coupling is a geometric consequence of the coplanar spiral magnetic moments. Although the conventional spin-orbit coupling in the superconducting substrate is crucial for stabilizing a non-collinear spin spiral ground state~\cite{PhysRevB.91.144424,Bruningetal2025}, the model we use does not account for it explicitly. Instead, it relies on the effective spin-orbit interaction generated by the spin spiral to realize topological superconductivity. Including the substrate spin-orbit coupling would only quantitatively modify the topological phase boundaries without changing the underlying topological phase.

\subsection{Topology}
The Hamiltonian in Eq.~\eqref{eq:gauge_transformed_H} is invariant under effective time-reversal symmetry given by the operator $T=\tau_0\sigma_x\mathcal{K}$, where $\mathcal{K}$ is complex conjugation with $T$ satisfying $T^2=1$. Combined with the particle-hole symmetry operator $C=\tau_y\sigma_y\mathcal{K}$, we have the chiral symmetry operator $S=CT=\tau_y\sigma_z$, which anticommutes with the Hamiltonian $\{S,H\}=0$. These symmetries place the system in the BDI topological class~\cite{PhysRevB.55.1142, PhysRevB.78.195125}. However, the BDI class does not have a topological ground state in 2D. Nevertheless, the Hamiltonian in Eq.~\eqref{eq:gauge_transformed_H} can be associated with a weak topological phase~\cite{PhysRevLett.98.106803, PhysRevB.75.121306, PhysRevB.79.195322}, in which the 2D system is viewed as a composition of parallel 1D BDI wires characterized by a $\mathbb{Z}$ invariant~\cite{PhysRevLett.89.077002, Ryu2010TenfoldWay}. The chiral symmetry then protects the multiple zero-energy Majorana states (one per wire) on the same edge from hybridizing with each other~\cite{sedlmayr2015majoranas}. This description requires lattice translation symmetry along the edges with the Majorana states. Therefore, even though each 1D BDI wire can host Majorana edge states, both translation and chiral symmetries are required to enable the corresponding 2D system to support Majorana flat bands, see Figs.~\ref{fig:bwo_pc_representative_points}\textcolor{blue}{(d,h)} for plots of the Majorana flat band. 

To compute the relevant topological number, we perform a Fourier transform of the Hamiltonian in Eq.~\eqref{eq:gauge_transformed_H} to obtain
\begin{align}\label{eq:Fourier_transform_H}
    H(\mathbf{k}) &= \sum_{k} \Psi_k^{\dagger}\big[\left(\mathcal{E}(\mathbf{k})-\mu\right)\,\tau_z\sigma_0+B\,\tau_0\sigma_x \nonumber \\
    & \hspace{2.2cm}-\Delta_{\rm bulk} \,\tau_x\sigma_0 + f(\mathbf{k})\tau_z\sigma_z\big]\Psi_k\,,
\end{align}
where $\mathbf{k} = (k_x,k_y)$, $\mathcal{E}(\mathbf{k}) = - t\,\{\cos(\pi q_x)\cos(k_x) + \cos(\pi q_y)\cos(k_y)\}$, and $f(\mathbf{k}) = t\,\{\sin(\pi q_x)\sin(k_x) + \sin(\pi q_y)\sin(k_y)\}$. Here, $\Delta_{\rm bulk}$ is the superconducting gap computed self-consistently in $k$-space by Fourier transforming Eq.~\eqref{eq:self_consistent_condition}. For a 1D BDI wire, the topological invariant is an integer winding number~\cite{WenZee1989}. For a 2D lattice with Majorana flat bands along the $y$-direction, see Fig.~\ref{fig:2D_lattice_schematic}, we can define a $k_y$-dependent winding number \cite{PhysRevB.83.224511, PhysRevB.85.035110, PhysRevLett.109.150408} as (see Appendix~\ref{app:winding_number} for details)
\begin{align}\label{eq:W_ky_formula}
    W(k_y) &= -\frac{1}{2}\sum_{f(\mathbf{k})=0}\nonumber \mathrm{sgn}\left[\Delta_{\rm bulk}\,\partial_{k_x}f(\mathbf{k})\right] \\
    &\hspace{0.2cm} \times \mathrm{sgn}\left[B^2-\left[\mathcal{E}(\mathbf{k})-\mu\right]^2-\Delta_{\rm bulk}^2+f^2(\mathbf{k})\right],
\end{align}
where the summation is taken for all $k_x$ satisfying the condition $f(\mathbf{k})=0$. In 1D wires, the spiral pitch is crucial for tuning the system into a topological phase~\cite{PhysRevB.88.020407}. Consequently, in a 2D system the spiral pitch vector $\mathbf{q}=(q_x,q_y)$ influences the number of Majorana states and their stability~\cite{sedlmayr_flat_2015,sedlmayr2015majoranas}. In this work, we consider $\mathbf{q}=(q_x,0)$, as a finite $q_y$ component has been found to be detrimental to Majorana flat bands along the $y$-direction~\cite{sedlmayr_flat_2015}.

\subsection{Self-consistent calculation of superconductivity}
To compute the superconducting order parameter $\Delta_r$ self-consistently, we solve the BdG Hamiltonian in Eq.~\eqref{eq:gauge_transformed_H} numerically using an iterative scheme on a grid of \mbox{$N_x\times N_y$} lattice sites. The system is modeled with a cylindrical geometry with open boundary conditions at sites $x=1,N_x$ and periodic boundary conditions along the $y$-direction. Starting from an initial guess for $\Delta_r$, the Hamiltonian in Eq.~\eqref{eq:gauge_transformed_H} is diagonalized (in the $\ell$-th iteration) to determine the new order parameter $\Delta_r^{(\ell)}$ via Eq.~\eqref{eq:self_consistent_condition}. In a standard self-consistent scheme, this updated order parameter yields the next iteration as $\Delta_r^{(\ell+1)} = \mathcal{T}(\Delta_r^{(\ell)})$, where $\mathcal{T}$ encapsulates Hamiltonian reconstruction, diagonalization, and gap recalculation. Convergence is achieved when the maximum change in $\Delta_r$ at each site $r$ is below a numerical threshold $\epsilon$ between successive iterations. To accelerate this convergence, we employ the Barzilai-Borwein method~\cite{Barzilai1988}. This method introduces an adaptive mixing factor $\alpha \in (0.001, 100]$~\cite{Holmvall2023SuperConga} to update the new order parameter as 
\begin{align}
   \Delta_r^{(\ell+1)} = (1-\alpha_{\ell+1})\Delta_r^{(\ell)} + \alpha_{\ell+1} \mathcal{T} (\Delta_r^{(\ell)})\,.   
\end{align}
Here, the mixing factor is updated as $\alpha_{\ell+1} = \alpha_\ell s_\ell$, where $s_\ell = s_\ell^{\mathrm{odd}}$ for odd $\ell$ and $s_\ell = s_\ell^{\mathrm{even}}$ for even $\ell$. These scaling factors depend on the real vector residual $\boldsymbol{\delta}_\ell = \Gamma[\mathcal{T}(\Delta_r^{(\ell)})] - \Gamma[\Delta_r^{(\ell)}]$, where $\Gamma[\bullet]$ stacks the real and imaginary parts $\big(\mathrm{Re}[\bullet], \mathrm{Im}[\bullet]\big)$. The scaling factors are determined by~\cite{Holmvall2023SuperConga}
\begin{align}
    s_\ell^{\mathrm{odd}} = \frac{\|\boldsymbol{\delta}_{\ell-1}\|_2^2}{\left| \boldsymbol{\delta}_{\ell-1} \cdot (\boldsymbol{\delta}_\ell - \boldsymbol{\delta}_{\ell-1}) \right|},~s_\ell^{\mathrm{even}} = \frac{\left| \boldsymbol{\delta}_{\ell-1} \cdot (\boldsymbol{\delta}_\ell - \boldsymbol{\delta}_{\ell-1}) \right|}{\|\boldsymbol{\delta}_\ell - \boldsymbol{\delta}_{\ell-1}\|_2^2}\,.
\end{align}

Self-consistent calculations for each parameter choice were performed with at least three distinct initial guesses of $\Delta_r$. First, we utilize a starting order parameter guess with edge-localized, site-to-site staggered amplitude modulation, and vanishing phase variation as an initial guess for the pair density wave phase. Second, the starting configurations for the phase crystal instead have a uniform amplitude with the ansatz for the phase variation being \mbox{$\theta(x,y) \propto [1+x/x_0]e^{-x/x_0}\sin(2\pi n y /N_y)$}, adopted from Ref.~\cite{PhysRevResearch.2.013104}. Here, we treat $x_0$ and $n$ as free parameters that govern the phase decay perpendicular to the edge ($x$-direction) and the phase modulation frequency along the edge ($y$-direction), respectively, and we try different frequencies in our initial guesses. Third, we also try uniform solutions. We employ these three unique types of start guesses over the full phase diagram to not bias the outcome. 
The converged solution with the lowest free energy is then assigned as the ground state. We compute the free energy using $\mathcal{F} = \langle H \rangle - T\mathcal{S}$, where $\langle H \rangle$ corresponds to the internal energy of the Hamiltonian in Eq.~\eqref{eq:gauge_transformed_H}, and $\mathcal{S}$ is the entropy expressed using the Fermi-Dirac function $f_n$ as $\mathcal{S} = -k_B\sum_n [f_n\, \ln(f_n) + (1-f_n)\,\ln(1-f_n)]$~\cite{Leggett_free_energy_1998}.

\subsection{Parameter choices}
All energies and temperatures are expressed in units of $t$, as we set $k_B=1$. For other parameters, we choose $(N_x,N_y) = (120,46),\mathbf{q}=(0.3,0),V=2t,$ and $\epsilon=10^{-6}$ and compute phase diagrams as functions of chemical potential $\mu$, magnetic field $B$, and temperature $T$. For a fixed spiral pitch $\mathbf{q}$, $\mu$ and $B$ govern the winding number in  Eq.~\eqref{eq:W_ky_formula}, which determines the number of Majorana edge states.
We find that this specific pitch value $\mathbf{q}$ supports the topological phase over a wide range of $\mu$ and $B$. A large $N_x$ is chosen to ensure that Majorana states localized at opposite edges of the system remain spatially isolated~\cite{Kim2018}, suppressing finite-size energy splitting caused by wavefunction overlap~\cite{Schneider2022} between different edges. In fact, in the topologically non-trivial regime, the bulk gap calculated in the finite-size cylinder matches $\Delta_{\rm bulk}$ within a $5\%$ relative deviation, indicating that the chosen geometry captures the bulk limit.
We note the choice of $V$ produces a relatively large superconducting order parameter $\Delta$, needed in order to keep the lattice size manageable and still capture the nonuniform superconducting states. Still, since we relate all results to topology and general features, we do not expect any qualitative differences in the presented results for other parameter choices, provided that the system is well within the topological phase hosting Majorana flat bands.

\section{Zero-temperature Phases}\label{sec:Phase_Diagrams}
\begin{figure}[!t]
    \centering
    \includegraphics[width=1\linewidth]{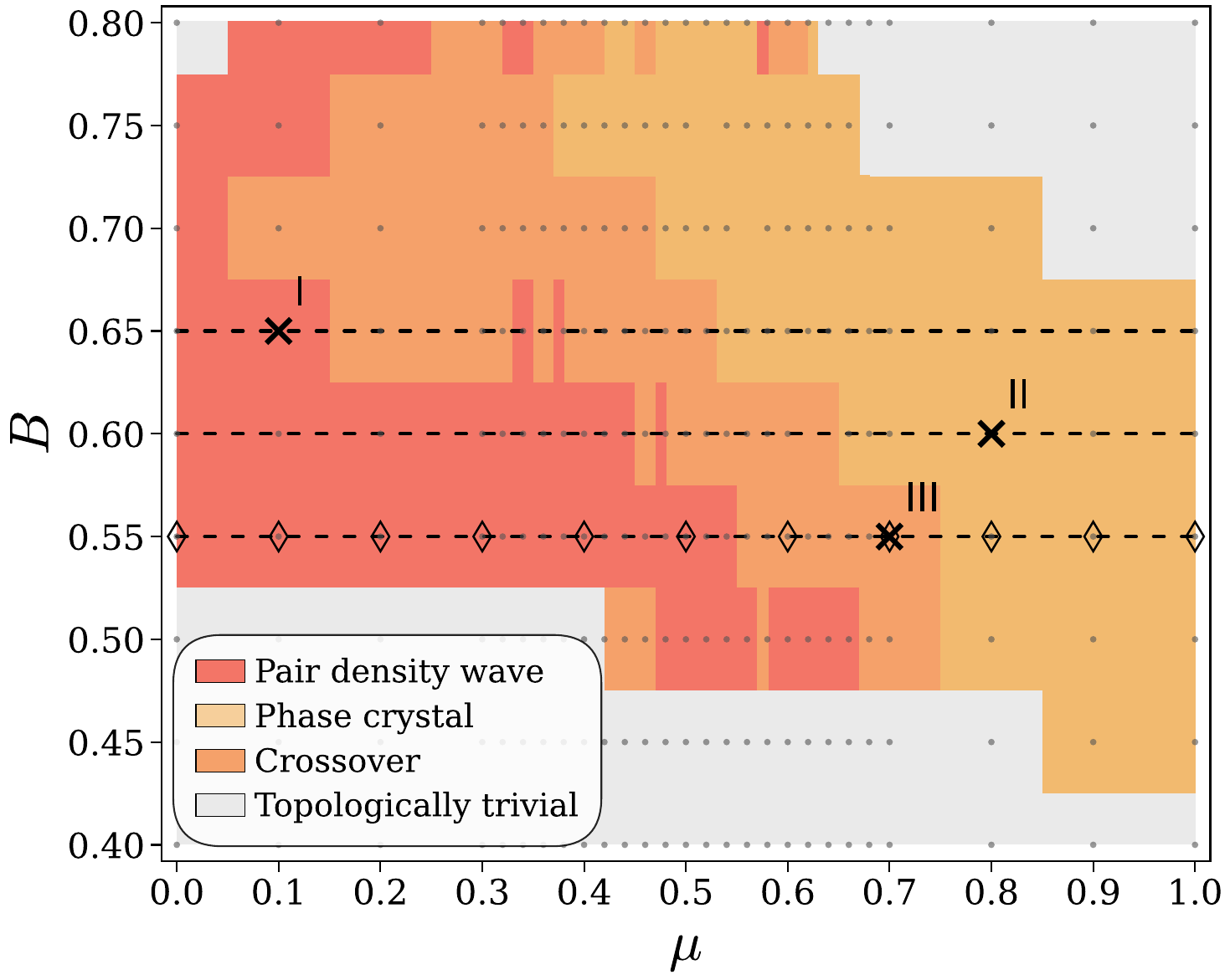}
    \caption{Ground-state phase diagram as a function of chemical potential $\mu$ and magnetic field $B$ at zero temperature. Gray dots indicate the calculated discrete data points. Crosses I, II, and III mark the representative $(\mu, B)$ parameters used for discussing the different nonuniform phases throughout Sec.~\ref{sec:Phase_Diagrams}. Dashed lines indicate values of $B$ used to detail amplitude and phase modulations in Sec.~\ref{sec:Transition_region}.   
    Open diamond markers indicate the value of $B$ used in finite temperature calculations in Sec.~\ref{sec:Finite_temperature_phase_diagram}. Close to the topological phase transitions, an energy landscape riddled with local minima makes it difficult to characterize the true ground state, leading to few point-to-point phase jumps along these boundaries.} \label{fig:zero_temp_phase_diagram}
\end{figure}

We first investigate the superconducting phase diagram as a function of chemical potential $\mu$ and magnetic field $B$ at zero temperature. The resulting zero-temperature phase diagram is shown in Fig.~\ref{fig:zero_temp_phase_diagram}. 
The red region corresponds to a pair density wave~\cite{PhysRevB.79.064515,Agterberg2008,pdw_footnote} phase, characterized by clear edge amplitude modulations but no phase variation. This phase breaks lattice translation symmetry along the edges, leading to the hybridization of Majorana states, as we discuss in Sec.~\ref{sec:Bond_wave_order}. Increasing the chemical potential results in a phase crystal~\cite{PhysRevResearch.2.013104}, denoted by the golden region in the phase diagram. This region instead has dominant phase modulations along the edge with minor accompanying amplitude modulations, as discussed in detail in Sec.~\ref{sec:Phase_crystal}. 
The phase crystal breaks chiral symmetry in addition to lattice translation symmetry. Although these two distinct nonuniform phases both differ in the symmetries they break, they share a common feature of lowering the free energy of the system by shifting the Majorana flat bands to finite energies. 
Furthermore, we find a crossover between these two nonuniform phases, shown in orange in the phase diagram. Within this region, we find the ground state to have both finite phase modulations and notable amplitude modulations along the edges.
Similar to a phase crystal, this crossover region breaks both translation and chiral symmetries. We detail the characterization of this regime in Sec.~\ref{sec:Transition_region}. Notably, the whole phase diagram is either in the topologically trivial uniform phase (gray) with no Majorana edge states, or in the pair density wave, phase crystal, or their crossover at zero temperature, i.e.,~we find no regime where the Majorana flat band survives, thanks to the superconducting order parameter adopting a highly inhomogeneous pattern near the edges. 
\subsection{Pair density wave}\label{sec:Bond_wave_order}
\begin{figure*}[htb]
    \centering
    \includegraphics[width=1\linewidth]{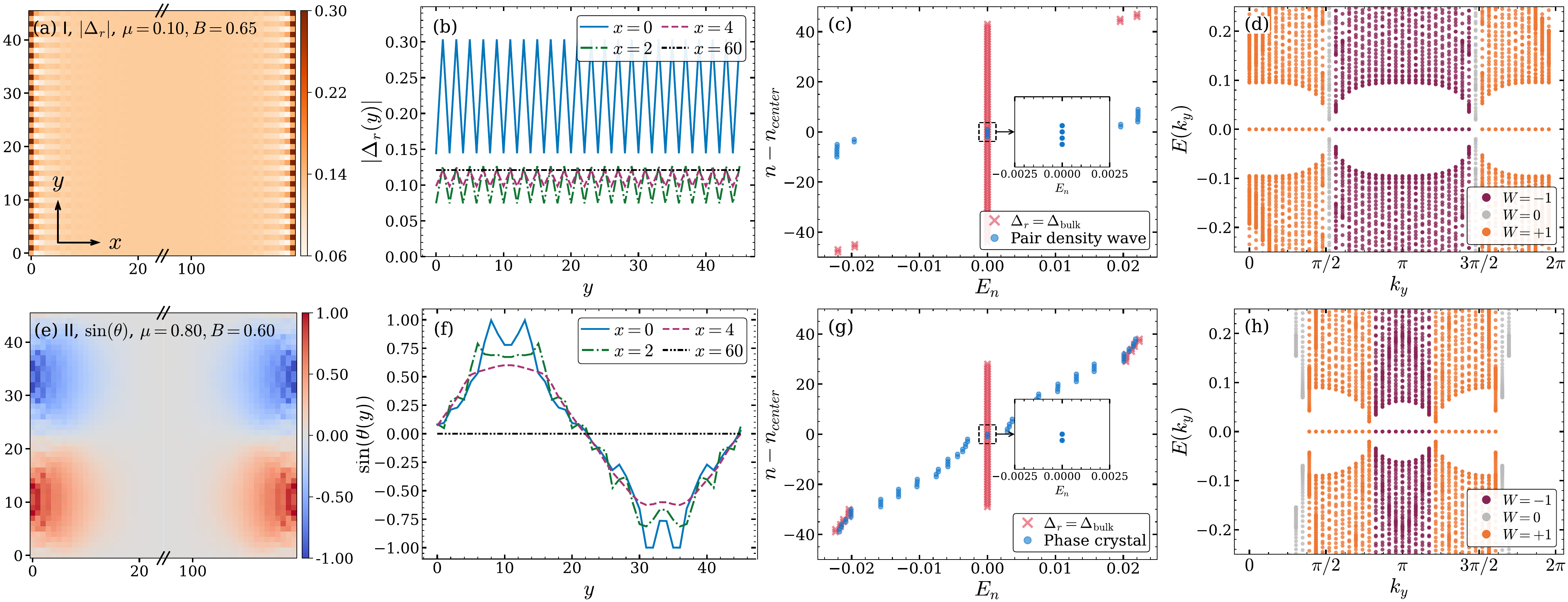}
    \caption{Pair density wave state [top row, (a-d)] and phase crystal state [bottom row, (e-h)], corresponding to points I at $(\mu, B) = (0.10, 0.65)$ and II at $(\mu, B) = (0.80, 0.60)$ in Fig.~\ref{fig:zero_temp_phase_diagram}, respectively. (a,e) Color density maps of the superconducting order parameter amplitude $|\Delta_r|$ at I and phase $\sin\theta$ at II, respectively. (b,f) Corresponding line plots of the amplitude and the phase modulations as a function of position $y$ along the edge for several fixed values of $x$ with $x=0$ being the edge and $x=60$ representing the bulk. (c,g) Low-energy eigenvalues $E_n$ as a function of eigenindex $n$ in pair density wave and phase crystal states (blue dots), compared to the uniform state (red crosses) with $\Delta_r=\Delta_{\rm bulk}$. (d,h) Low-energy band structure in a semi-infinite geometry, obtained by a Fourier transform along $y$ with $\Delta_r=\Delta_{\rm bulk}$ as a function of $k_y$. Colors denote the winding number $W(k_y)$.}     
    \label{fig:bwo_pc_representative_points}
\end{figure*}

We start by examining the properties of the pair density wave, choosing the representative parameters marked by cross I ($\mu = 0.10$ and $B = 0.65$) in Fig.~\ref{fig:zero_temp_phase_diagram}. In this state, the amplitude of the superconducting order parameter $|\Delta_r|$ oscillates rapidly along the edges with a large wave vector $Q_y$, while remaining uniform in the bulk, see Fig.~\ref{fig:bwo_pc_representative_points}\textcolor{blue}{(a)}. However, its phase $\theta_r$ remains constant with no spatial variation (see Appendix~\ref{app:additional_bwo_pc_data}). In Fig.~\ref{fig:bwo_pc_representative_points}\textcolor{blue}{(b)} we extract the amplitude $|\Delta_r|$ along the $y$-direction in regions near the edge ($x=0,2,4$) and also in the bulk ($x=60$). This shows a pair density wave characterized by site-to-site staggered edge amplitude modulations, this is thus a bond-wave order~\cite{wang_fate_2014}. These amplitude modulations break the translational invariance along the edges of the system, but the chiral symmetry is still preserved, as the lack of phase modulations in the pair density wave leaves the time-reversal symmetry ($T^2=1$) unbroken. The finite transverse extent of these amplitude modulations into the bulk also indicates that this is not a strictly one-dimensional phenomenon. The broken translation symmetry is detrimental to the Majorana flat band due to its weak topological protection.
In Fig.~\ref{fig:bwo_pc_representative_points}\textcolor{blue}{(c)} we illustrate this by plotting the distribution of low-energy eigenvalues $E_n$ for both the pair density wave (blue circles) and the uniform superconducting state (red crosses). The latter is obtained by diagonalizing the Hamiltonian in Eq.~\eqref{eq:gauge_transformed_H} with a fixed $\Delta_r = \Delta_{\rm bulk}$, the self-consistent bulk superconducting gap. We find that the pair density wave strongly hybridizes most of the Majorana flat band states such that they are removed from zero energy. This lowers the free energy of the system, and, since the pair density wave is the ground state for a large range of parameters, the energetic cost of the nonuniform state is clearly outweighed by the gain in free energy from removing these Majorana states.

To understand why the pair density wave forms and also favors the bond-wave order configuration, we perform a Fourier transform of the Hamiltonian in Eq.~\eqref{eq:gauge_transformed_H} along $y$ with an open boundary condition along $x$. This results in a set of independent wires of length $N_x$ labeled by momenta $k_y = 2\pi n/N_y$, with $n=0,1,\dots, N_y-1$. Each $k_y$-resolved wire can enter the topological phase to host Majorana edge states~\cite{PhysRevB.89.174514,sedlmayr_flat_2015}, and their band-structure with uniform $\Delta_r = \Delta_{\rm bulk}$ is plotted in Fig.~\ref{fig:bwo_pc_representative_points}\textcolor{blue}{(d)}. The colors denote the winding number $W(k_y)$ calculated from Eq.~\eqref{eq:W_ky_formula}. This indicates that the Majorana edge states correspond to $W(k_y)=\pm 1$ and terminate at the gap-closing points with a trivial winding $W(k_y) = 0$~\cite{PhysRevB.83.224511,PhysRevB.86.104509,Yuan2014}. Here, chiral symmetry protects the Majorana states with identical windings from coupling with each other~\cite{sedlmayr_flat_2015}. However, states on the same edge with opposite windings are still allowed to hybridize~\cite{MizushimaSato2013, wang_fate_2014}. Such a coupling requires scattering between different $k_y$-resolved wires of opposite windings. The pair density wave exactly enables this scattering by inducing edge-localized amplitude modulations. These modulations have a dominant wave vector $Q_y$, which corresponds to the momentum difference between the $k_y$-resolved wires with opposite windings. 

In Fig.~\ref{fig:bwo_pc_representative_points}\textcolor{blue}{(d)}, the wires with $W(k_y)=+1$ can be efficiently coupled to wires with opposite winding if they are scattered by $Q_y=\pi$. This process is enabled by the bond wave order configuration of the pair density wave, which doubles the unit cell in real-space through its site-to-site staggered modulation of $|\Delta_r|$. The resulting unit cell doubling halves the Brillouin zone, which folds the Majorana flat band onto itself. Consequently, the Majorana states with opposite windings hybridize and open a single uniform gap as shown in Fig.~\ref{fig:bwo_pc_representative_points}\textcolor{blue}{(c)}. Notably, it is sufficient to modulate the amplitude along the edges, while keeping it constant in the bulk to hybridize the Majorana states, which minimizes the condensation energy cost of amplitude modulations. 
Our calculations show that the bond-wave order configuration of the pair density wave is largely maintained for $\mu \lessapprox 0.2$.
As $|\mu|$ increases, the number of $k_y$-resolved wires with trivial winding $W=0$ increases, and folding the Brillouin zone in half is now not energetically sufficient to hybridize the Majorana flat band.  The system then instead scatters non-trivial $k_y$-resolved wires at different values of $Q_y$ rather than a single dominant $Q_y=\pi$. As a result, the pair density wave changes from bond-wave order to a configuration that exhibits a combination of multiple modulation wave vectors, albeit all with short wavelengths. This causes multiple overlapping band foldings in the Brillouin zone. The resulting hybridization of the Majorana flat band leads to sub-gap states with numerous tiny gap openings. 

As the pair density wave relies on coupling Majorana states on the same edge with opposite windings~\cite{wang_fate_2014}, the remnant unhybridized states correspond to the difference $2\left|W_+ - W_-\right|$. Here, $W_{\pm}$ denotes the number of $\pm1$ winding numbers in the $k_y$-resolved band-structure. The remnant Majorana states remain at zero energy protected by the chiral symmetry. For the representative parameters $(\mu, B) = (0.10,0.65)$ in Fig.~\ref{fig:bwo_pc_representative_points}\textcolor{blue}{(a-d)}, $W_+=21$ and $W_- = 23$, which leaves four Majorana states at zero-energy, in agreement with the inset of Fig.~\ref{fig:bwo_pc_representative_points}\textcolor{blue}{(c)}. Generally, in the uniform superconducting state, the maximum number of Majorana states well-within the topological phase region is obtained at $\mu=0$. At this value of $\mu$ and for an even $N_y$, the difference between $W_+$ and $W_-$ is zero, and the pair density wave thus gaps out all of the Majorana states. In contrast, this difference is one for an odd $N_y$, leaving a pair of Majorana states at zero energy. To summarize, the pair density wave is driven by the presence of Majorana edge states with different winding numbers. It breaks lattice translation symmetry through edge-localized amplitude modulations such that it gaps out all possible Majorana edge states with opposite winding numbers. We note that this is different from conventional finite-momentum pairing, or FFLO pairing, ~\cite{PhysRev.135.A550,Larkin:1964wok}, which is an intrinsic bulk effect or its remnant at a surface~\cite{PhysRevLett.122.165302}, found across the superconductor-normal transition and primarily driven by spin splitting of the Fermi surface.  


\subsection{Phase crystal}\label{sec:Phase_crystal}
We next investigate the characteristics of the phase crystal for the parameters marked by the cross II (\mbox{$\mu = 0.80$} and $B = 0.60$) in Fig.~\ref{fig:zero_temp_phase_diagram} and report the details in the bottom panels in Fig.~\ref{fig:bwo_pc_representative_points}. As shown in Figs.~\ref{fig:bwo_pc_representative_points}\textcolor{blue}{(e,f)} in this state, the phase $\theta_r$ of the superconducting order parameter modulates along the edges but stays uniform in the bulk. The modulation frequency is set by the decay length of the zero-energy states into the bulk, both governed by the superconducting coherence length~\cite{PhysRevResearch.2.013104}. Driven by these strong phase modulations, the amplitude $|\Delta_r|$ also acquires a modest spatial variation along the edges that is much weaker and smoother compared to the pair density wave. Importantly, the modulation length scale of these amplitude modulations is set by the phase modulations. Specifically, the amplitude variation peaks when the phase gradient is zero (see Appendix~\ref{app:additional_bwo_pc_data} for details). The spatially varying phase results in spontaneous superfluid momentum, $\mathbf{p}_{\mathrm{S}} \propto \nabla\theta_r$, which non-locally drive staggered loop currents (see Appendix~\ref{app:additional_bwo_pc_data}) that break both translation and time-reversal symmetries~\cite{wang_fate_2014,Hakansson2015, Holmvall2018,PhysRevB.99.184511, PhysRevResearch.2.013104, PhysRevResearch.2.043198, Chakraborty2022,PhysRevB.110.064502, PhysRevB.109.L180509}. The spontaneous superfluid momentum shifts the Majorana flat band to finite energies, as shown in Fig.~\ref{fig:bwo_pc_representative_points}\textcolor{blue}{(g)}. Formally such an energy shift is allowed because the phase crystal breaks the chiral symmetry, and without this symmetry, the Majorana states on the same edge may hybridize with each other and shift to finite energies~\cite{sedlmayr2015majoranas}. Although the phase modulations and associated loop currents would increase the free energy of the condensate in the bulk, the existence of zero-energy edge states leads to effectively negative and non-local correlations in the superfluid stiffness tensor~\cite{PhysRevResearch.2.013104}. This instead favors edge-localized phase gradients that lower the overall free energy of the system. The resulting free energy gain corresponds to shifting the Majorana states away from zero energy.

To understand why the system favors a phase crystal over a pair density wave, we plot the $k_y$-resolved band structure of the uniform state with $\Delta_r=\Delta_{\rm bulk}$, in Fig.~\ref{fig:bwo_pc_representative_points}\textcolor{blue}{(h)}. This shows that the flat band of Majorana states is dominated by one non-trivial winding, $W_+ \gg W_-$. As the pair density wave relies on scattering between opposite windings, this large imbalance results in many remnant Majorana states. As a consequence, the pair density wave is no longer energetically efficient in hybridizing the Majorana flat bands. Instead, the system forms a phase crystal that spontaneously breaks the chiral symmetry and thereby hybridizes the Majorana states, lowering the free energy. The periodicity of the resulting phase crystal is governed by a delicate balance between the energy gain of phase variations close to the edge due to effectively negative stiffness and the cost of phase variations further from the edge~\cite{PhysRevResearch.2.013104}. This balance is also known to be influenced by superconducting coherence length and other relevant system parameters such as finite size~\cite{PhysRevB.99.184511,PhysRevResearch.2.043198,PhysRevB.111.094513}. For instance, at $(\mu, B) = (0.80,0.50)$ and $(0.90,0.55)$, we obtain a phase crystal instead with six and four phase gradient nodes along $y$, respectively, as the ground state of the system (see Appendix~\ref{app:additional_bwo_pc_data} for details).

The absence of chiral symmetry for the phase crystal facilitates a pair-wise hybridization of the Majorana states along the same edge. Consequently, the system can only host a maximum of two Majorana states, with one localized on each edge, depending on the parity of the total number of non-trivial $k_y$-resolved wires, given by $2\left[(W_+ + W_-)\pmod 2\right]$. An even number of non-trivial wires leads to an even number of Majorana states along a single edge, which hybridize and gap out completely, while an odd number of non-trivial wires results in one unpaired state surviving on each edge. For the representative parameters $(\mu,B) = (0.80,0.60)$, $W_+ + W_- = 11$, which leaves a pair of Majorana states, see inset in Fig.~\ref{fig:bwo_pc_representative_points}\textcolor{blue}{(g)}. A similar odd-even effect arising from hybridization of multiple Majorana bound states has also been reported in Refs.~\cite{wang_fate_2014, MizushimaSato2013}. Well within the topological region, we observe that at $\mu=1.0$, the sum of $W_+$ and $W_-$ is even, causing the phase crystal to gap out all Majorana states, regardless of whether $N_y$ is odd or even. To summarize, a phase crystal emerges when the amplitude modulations of a pair density wave are insufficient to hybridize the Majorana flat bands due to a mismatch in the winding numbers. Despite the fact that spontaneous phase modulations and currents usually cost energy in a bulk superconductor, the presence of Majorana edge states at zero-energy leads to effectively negative superfluid stiffness correlations~\cite{PhysRevResearch.2.013104} such that the edge phase modulations instead minimize the free energy by breaking both translation and chiral symmetries. Hence, weak topology and the winding numbers of Majorana states strongly influence the type of emergent nonuniform superconducting phase. 
\begin{figure*}[t!]
    \centering
    \includegraphics[width=1\linewidth]{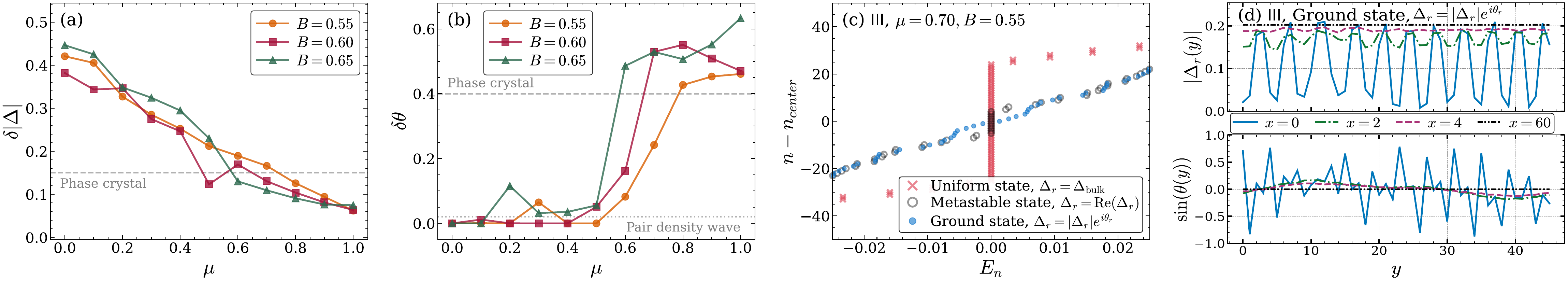}
    \caption{(a,b) Relative change in the superconducting order parameter amplitude $\delta|\Delta|$ and its phase $\delta\theta$, respectively, plotted as a function of chemical potential $\mu$ for representative values of $B$ indicated by dashed lines in Fig.~\ref{fig:zero_temp_phase_diagram}. (c) Low-energy eigenvalues $E_n$ plotted as a function of eigenindex $n$ corresponding to point III at $(\mu, B) = (0.70, 0.55)$ in Fig.~\ref{fig:zero_temp_phase_diagram}. The ground state (blue dots) is compared to the uniform state (red crosses) with $\Delta_r=\Delta_{\rm bulk}$, and the metastable pair density wave (gray circles) obtained via self-consistent calculations enforcing a real-valued superconducting order parameter. (d) Line plots showing the amplitude modulations (upper panel) and the phase modulations (lower panel) of the ground state at point III as a function of position $y$ along the edge for several fixed values of $x$ with $x=0$ being the edge and $x=60$ representing the bulk.}     
    \label{fig:transition_region_points}
\end{figure*}

\subsection{Crossover between pair density wave and phase crystal}\label{sec:Transition_region}

We finally discuss the crossover (indicated by the orange area in Fig.~\ref{fig:zero_temp_phase_diagram}) from the pair density wave to the phase crystal, driven primarily by increasing $|\mu|$. This region hosts states that inherit characteristics from both superconducting phases. Note that the pair density wave and the phase crystal are distinct phases: they break different symmetries, use different mechanisms to gap out the Majorana flat band to lower the free energy, and have different emergent properties. There is still a crossover region between them that mixes these characteristics.
Specifically, the crossover region shows edge-localized modulations of the amplitude $|\Delta_r|$ and phase $\theta_r$, which are weaker than those in the pair density wave and the phase crystal, respectively. To formally characterize this crossover, we investigate the relative changes in the order parameter amplitude $\delta|\Delta|$ and phase $\delta\theta$ along the edge. These quantities are calculated as the spatial average of the root-mean-square (RMS) deviations along $y$ given by \mbox{$\delta\zeta(\mu, B) = \frac{1}{N}\sum_{x=0}^{N-1} \langle \zeta(x) -\overline{\zeta}(x)\rangle_{\mathrm{rms},y}$}, where $\overline{\zeta}$ denotes the arithmetic mean. Here, $\zeta$ denotes the normalized amplitude $|\Delta_r|_{\mathrm{norm}} = |\Delta_r|/\Delta_{\rm bulk}$ or  $\sin\theta$. We set the averaging window along $x$ to $N=3$, as it is sufficient to capture the dominant amplitude and phase modulations. We plot $\delta|\Delta|$ and $\delta\theta$ in Figs.~\ref{fig:transition_region_points}\textcolor{blue}{(a,b)}, respectively, as functions of $|\mu|$ for representative values of $B$ indicated by dashed lines in Fig.~\ref{fig:zero_temp_phase_diagram}.  

In Figs.~\ref{fig:transition_region_points}\textcolor{blue}{(a,b)} the discrete data points exhibiting finite amplitude variation $|\delta\Delta| \neq 0$ alongside vanishing phase variations $\delta\theta \approx 0$ characterize the pair density wave. This condition is strictly tied to the topological constraint of nearly balanced opposite windings $W_\pm$, which enables dominant amplitude modulations to hybridize the Majorana flat band, as explained in Sec.~\ref{sec:Bond_wave_order}. Consequently, any parameter regime satisfying this topological constraint will result in a pair density wave with $\delta\theta \approx 0$, corresponding to the dotted gray line in Fig.~\ref{fig:transition_region_points}\textcolor{blue}{(b)}.
In contrast, the phase crystal, defined in Sec.~\ref{sec:Phase_crystal}, exhibits strong phase modulations that drive smooth amplitude variations along the edge, such that $\delta\theta > \delta|\Delta|$. Because the pair density wave cannot efficiently hybridize a Majorana flat band with large imbalance between opposite windings $W_\pm \gg W_\mp$, the phase crystal emerges in this parameter regime to lower the free energy of the system. 
However, due to the crossover between the two nonuniform states, defining a quantitative boundary for dominant phase modulations depends on the parameter regime. Nonetheless, the qualitative features of a phase crystal remain unchanged~\cite{PhysRevResearch.2.013104}. For our choice of parameters, we define the phase crystal by \mbox{$\delta\theta > 0.4$} (representing $40\%$ of the maximum phase \mbox{$|\sin\theta|_{\mathrm{max}}=1$}), while the corresponding amplitude variations $\delta|\Delta|$ remain roughly below $0.15$ (representing $15\%$ of the normalized bulk value $|\Delta_{\rm bulk}|_{\mathrm{norm}}=1$), see gray dashed lines in Figs.~\ref{fig:transition_region_points}\textcolor{blue}{(a,b)}. Given these criteria, we can then define the crossover as the region where the relative change in phase modulations remains below $40\%$ and the amplitude modulations roughly exceed $15\%$ of the normalized bulk gap~\cite{crossover_footnote}. We note that any changes to the numerical bounds of the phase crystal, that align with its defining topological constraints, will only shift the crossover boundary internally, without any qualitative changes to the overall phase diagram.
\begin{figure*}[htb]
    \centering
    \includegraphics[width=1\linewidth]{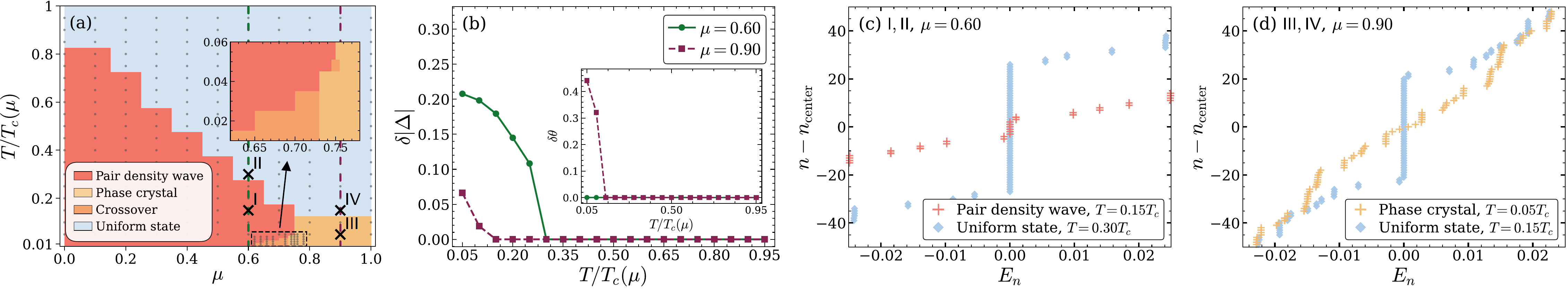}
    \caption{(a) Ground-state phase diagram as a function of chemical potential $\mu$ and temperature $T$ at $B=0.55$. Here, $T$ is normalized by the bulk superconducting transition temperature $T_c(\mu)$. Gray dots indicate the calculated discrete data points. The inset shows the small area occupied by the crossover region between pair density wave and phase crystal. (b) Relative change in the superconducting order parameter amplitude $\delta|\Delta|$ and its phase $\delta\theta$ (inset) plotted as a function of normalized temperature $T/T_c(\mu)$ for fixed values of chemical potential $\mu$ indicated by the dashed lines in (a). Panels (c), (d) show low-energy eigenvalues $E_n$ plotted as a function of eigenindex $n$ for $\mu=0.60, T=0.15T_c, 0.30T_c$ [I, II in (a)] and $\mu=0.90, T=0.05T_c, 0.15T_c$ [III, IV in (a)], respectively.}
    \label{fig:finite_temperature_points}
\end{figure*}

We attribute the existence of the crossover region to remnant Majorana states with identical windings that cannot hybridize within the pure pair density wave state. Typically, the imbalance between $W_+$ and $W_-$ increases with increasing $|\mu|$. Here, the system develops finite phase modulations in addition to existing amplitude modulations of the pair density wave to hybridize the remnant Majorana states. This further reduces the free energy of the system, resulting in a crossover where both amplitude and phase modulations coexist on a rather equal footing. To confirm this, we performed additional self-consistent calculations in the crossover region by restricting the order parameter to be real valued, allowing only for amplitude variations. These calculations result in a metastable pair density wave with higher free energy compared to the ground state. The low-energy eigenvalues $E_n$ at the representative parameters $\mu = 0.70$ and $B=0.55$ marked by the cross III in Fig.~\ref{fig:zero_temp_phase_diagram}, are shown in Fig.~\ref{fig:transition_region_points}\textcolor{blue}{(c)}. This clearly shows how the real-valued metastable solution (gray circles) still hosts some Majorana states, although substantially reduced from the uniform $\Delta_r = \Delta_{\rm bulk}$ (red crosses). In contrast, the ground state of the system (blue dots) gaps out these remnant Majorana states with both amplitude and phase modulations, as shown in Fig.~\ref{fig:transition_region_points}\textcolor{blue}{(d)}, to further lower the free energy.

It has previously been interpreted~\cite{wang_fate_2014,sedlmayr_flat_2015} that pure amplitude modulations ($\delta\theta \approx 0$) appear only at $\mu=0$. These studies further suggested that, for increasing $N_y$, the system would generally prefer to gap out the Majorana flat bands through phase oscillations ($\delta\theta \neq 0$). However, we perform additional calculations on a $(N_x, N_y) = (60, 80)$ lattice and find that the zero-temperature phase diagram qualitatively remains unchanged even for larger $N_y$. Additional numerical checks, presented in Appendix~\ref{app:additional_lattice_size_data}, show that the pair density wave, phase crystal, and their crossover remain stable for larger $N_y$. Hence, our results clarify that for a weak topological superconductor in the BDI class, dominant amplitude modulations (pair density wave: $\delta|\Delta|>0, \delta\theta \approx 0$) and phase modulations (phase crystal: \mbox{$\delta\theta > \delta|\Delta|$}) represent two different nonuniform superconducting phases that emerge to reduce the free energy of the system, but through different mechanisms.

\section{Temperature Dependent Phase diagram}
\label{sec:Finite_temperature_phase_diagram}
Having studied the two nonuniform superconducting phases at zero temperature, we next investigate their finite-temperature behavior within the mean-field approximation. In particular, we show the ground-state phase diagram as a function of temperature and chemical potential $\mu$ in Fig.~\ref{fig:finite_temperature_points}\textcolor{blue}{(a)}. The temperature $T$ is normalized by $T_c(\mu)$, which is the bulk superconducting transition temperature. These results are calculated for the representative value \mbox{$B=0.55$}, indicated by the open diamond markers in Fig.~\ref{fig:zero_temp_phase_diagram}. Numerical checks at different $B$ show no qualitative changes in temperature behavior, beyond the phase crystal appearing as the ground state at lower values of $\mu$ at higher $B$. Unlike the zero-temperature limit, the finite-temperature phase diagram features the topologically non-trivial uniform phase with Majorana flat band as a self-consistent solution appearing at sufficiently high temperatures, while the crossover region between the pair density wave and the phase crystal substantially diminishes.  

In Fig.~\ref{fig:finite_temperature_points}\textcolor{blue}{(a)}, the boundary between the pair density wave and the (self-consistent) uniform state solution sets a transition temperature $T^{*}_{\text{PDW}}(\mu)$. This value decreases linearly from around $80\%$ of the bulk transition temperature at $\mu=0.0$ to around $10\%$ at $\mu\approx 0.75$. This behavior of $T^{*}_{\text{PDW}}(\mu)$ follows the monotonic weakening of the amplitude modulations $\delta|\Delta|$ with increasing $|\mu|$ as already seen in Fig.~\ref{fig:transition_region_points}\textcolor{blue}{(a)}. Furthermore, we compute the mean RMS deviations $\delta|\Delta|$ and $\delta\theta$ as a function of $T/T_c$ at $\mu = 0.60$, indicated by a dashed green line in Fig.~\ref{fig:finite_temperature_points}\textcolor{blue}{(a)}, to analyze the transition of the pair density wave to the uniform state solution in Fig.~\ref{fig:finite_temperature_points}\textcolor{blue}{(b)} (solid green line). Here, $\delta|\Delta|$ exhibits a monotonic decrease with increasing temperature, while $\delta\theta$ remains zero throughout the transition, as expected. In Fig.~\ref{fig:finite_temperature_points}\textcolor{blue}{(c)}, we also plot the distribution of low-energy eigenvalues $E_n$ at the same $\mu=0.60$ for the two temperatures corresponding to crosses I and II in Fig.~\ref{fig:finite_temperature_points}\textcolor{blue}{(a)}. Above $T>T^*_{\mathrm{PDW}}$ the system hosts the full Majorana flat band, but below $T^*_{\mathrm{PDW}}$ only the remnant Majorana states from the pair density wave survive. The transition temperature of the phase crystal $T^{*}_{\text{PC}}$ remains at $10\%$ of the bulk $T_c(\mu)$ throughout the phase diagram where it exists in Fig.~\ref{fig:finite_temperature_points}\textcolor{blue}{(a)}. However, depending on system parameters, e.g.,~in systems with an effectively shorter coherence length, the transition temperature $T^{*}_{\text{PC}}$ has previously been shown to be significantly enhanced above $20\text{--}30\%$ of the bulk $T_c$~\cite{PhysRevResearch.2.043198,Chakraborty2022}. The mean RMS deviations at $\mu=0.90$, indicated by the dashed maroon line in Fig.~\ref{fig:finite_temperature_points}\textcolor{blue}{(a)} are plotted in Fig.~\ref{fig:finite_temperature_points}\textcolor{blue}{(b)} and show that $\delta\theta$ decreases with increasing temperature. Since chiral symmetry remains broken throughout $T<T^*_{\mathrm{PC}}$, the Majorana flat band reappears only when $\delta\theta$ vanishes above $T^*_{\mathrm{PC}}$, as illustrated by the low-energy eigenvalues in  Fig.~\ref{fig:finite_temperature_points}\textcolor{blue}{(d)}, corresponding to the crosses III and IV in Fig.~\ref{fig:finite_temperature_points}\textcolor{blue}{(a)}. 

Furthermore, we find that the crossover between the pair density wave and the phase crystal occupies a considerably smaller area in the finite-temperature phase diagram, as shown in the inset of Fig.~\ref{fig:finite_temperature_points}\textcolor{blue}{(a)}. Our calculations show that the crossover region persists only up to approximately $4\text{--}5\%$ of the bulk $T_c(\mu)$. This, combined with the lower transition temperature of the phase crystal, suggests that phase modulations are more sensitive to thermal effects than the amplitude modulations. We can attribute this fragility at finite temperatures to at least two effects. First, as $|\mu|$ increases, the number of Majorana states decreases. This leads to a suppression of all emergent nonuniform phases. Second, the superconducting bulk gap $\Delta_{\rm bulk}$ decreases with increasing $|\mu|$ and $B$. At finite temperatures, this facilitates the thermal occupation of states above zero energy, thereby diminishing the free energy gain obtained by shifting the zero-energy edge states. As the phase $\theta_r$ modulated regime in the zero-temperature phase diagram occupies regions of relatively high $\mu$ and $B$, it thereby becomes more sensitive to temperature effects. \section{Summary and Outlook}\label{sec:Summary_Conclusions}
In this work we examine the mean-field thermodynamic instabilities of a two-dimensional Majorana flat band model towards inhomogeneous superconducting ground states. In particular, we study a tight-binding model of a helical spin chain on an $s$-wave superconductor with chiral symmetry, belonging to the BDI topological class~\cite{sedlmayr_flat_2015}. This system supports a weak topological phase~\cite{PhysRevLett.98.106803, PhysRevB.75.121306, PhysRevB.79.195322} in the form of a translationally invariant stack of 1D BDI wires, each characterized by a winding number. Earlier studies on such models have shown that pairing modulations in the superconducting order parameter were detrimental to the stability of Majorana flat bands~\cite{Li_2013,PhysRevB.89.174514,MizushimaSato2013,wang_fate_2014,PhysRevA.92.023621, PhysRevLett.121.185302}. We clarify that these modulations are two different symmetry-breaking nonuniform superconducting ground states, a pair density wave and a phase crystal, that minimize the free energy by removing the zero-energy Majorana states using different mechanisms. The pair density wave state breaks translational invariance via amplitude $|\Delta_r|$ modulations without any phase $\theta_r$ modulations. This state allows for hybridization of Majorana states with different winding numbers, thereby removing them from zero energy. In contrast, the phase crystal exhibits a spatially varying phase $\theta_r$ that breaks both translation and chiral symmetries, with only accompanied small amplitude modulations driven by the varying phase. The phase variations correspond to a finite superfluid momentum that shifts the Majorana states to finite energies and drives spontaneous currents. Overall, the competition between these nonuniform phases is governed by the winding numbers of the constituent 1D BDI wires, which primarily depend on the chemical potential. In the zero-temperature limit, we find that these two phases are bridged by a crossover region, where the order parameter has comparable modulations in both amplitude and phase, where Majorana states are effectively removed by both hybridization and shifted due to spontaneous superfluid momentum. 

At finite temperature, we find that the nonuniform phases exhibit distinct transition temperatures to the uniform state, hosting the Majorana flat band. Specifically, the pair density wave transition temperature decreases from $80\%$ to $10\%$ of $T_c$ as the chemical potential potential increases, a trend consistent with the weakening amplitude variations $\delta|\Delta|$ observed in the zero-temperature limit. The transition temperature for the phase crystal remains approximately $10\%$ of $T_c$ throughout the phase diagram. Based on previous studies, we expect that the phase crystal transition temperature can be substantially enhanced e.g.~in systems with an effectively shorter coherence length~\cite{PhysRevResearch.2.043198,Chakraborty2022}, making it significantly more competitive. The crossover region occupies a much smaller region in the finite-temperature phase diagram. We attribute the relative thermal instability of the phase modulations to a relatively lower number of Majorana states and a smaller superconducting bulk gap. 

The broader message in this work is that the Majorana flat bands in at least a weak topological superconductor are extremely unstable towards spatial modulations of the superconducting order parameter. In particular, we find that the order parameter easily modulates its amplitude $|\Delta_r|$ and also its phase $\theta_r$ when the amplitude modulations are insufficient to remove the Majorana flat band. Their dependence on winding numbers shows that the underlying topology dictates the nature of the emergent nonuniform superconducting phase. Therefore, systems with Majorana flat bands become a natural playground for strongly nonuniform superconducting states. 

A natural follow up of our work is investigating the effects of time-reversal invariant disorder. While such disorder keeps the system within the BDI class, leaving the Majorana flat bands intact~\cite{sedlmayr_flat_2015}, its impact on the emergent nonuniform superconducting ground states is yet to be studied. Additionally, our finite-temperature mean-field calculations do not account for possible additional effects due to thermal phase fluctuations, which needs further investigation. Moreover, our work only considers a weak topological system with lattice translation and chiral symmetries. Exploring systems across different topological classes with stronger symmetries protecting the Majorana flat bands not only broadens our understanding of the interplay between energy minimization and topology, but also provides insights into the stability of Majorana states. For example, it has been demonstrated that imposing mirror symmetry protects degenerate Majorana Kramers pairs against inhomogeneities of the order parameter in the $s$-wave superfluid belonging to the DIII topological class~\cite{PhysRevLett.121.185302}. Another complementary direction to assess the interplay between spontaneous symmetry-breaking, interactions, and topology might be provided by the generalized Landau paradigm~\cite{tmvy-vsqd}.


\begin{acknowledgments}
We acknowledge A.~Theiler and A.~K.~Ghosh for helpful discussions. We acknowledge financial support from the European Union through the European Research Council (ERC) under the European Union’s Horizon 2020 research and innovation programme (ERC-2022-CoG, Grant agreement No.~101087096).
Views and opinions expressed are, however, those of the authors only and do not necessarily reflect those of the European Union or the European Research Council Executive Agency. Neither the European Union nor the granting authority can be held responsible for them.
The computations were enabled by the Berzelius resource provided by the Knut and Alice Wallenberg Foundation at the National Supercomputer Centre.
Additional computations and data handling were enabled by resources provided by the National Academic Infrastructure for Supercomputing in Sweden (NAISS) at NSC, UPPMAX, PDC, and HPC2N, partially funded by the Swedish Research Council through grant agreements No.~2022-06725.
We further acknowledge NAISS for awarding this project access to the LUMI supercomputer, owned by the EuroHPC Joint Undertaking and hosted by CSC (Finland) and the LUMI consortium.
\end{acknowledgments}

\section{Data Availability}
Data supporting the findings of this article are openly available~\cite{Zenodo_data_and_code}.
\clearpage
\appendix


\onecolumngrid
\section{Gauge transformation introducing the effective spin-orbit coupling}
\label{app:gauge_transformation}
We apply a local gauge transformation $\Psi_r \rightarrow U_r\Psi_r$ to the Hamiltonian in Eq.~\eqref{eq:MFB_stability_true_H} using the unitary operator $U_r = \tau_0\,\text{exp}[i(\phi_r/2)\sigma_{z}]$ to obtain
\begin{align}\label{app_eq:MFB_stability_UdaggerHU}
    H = \sum_{r} \Psi_r^{\dagger}\left[-\mu\,\tau_z\sigma_0+B\,\tau_0\sigma_x + \left(-\Delta_r \,\frac{1}{2}(\tau_x+i\tau_y)\sigma_0 + \text{H.c.}\right)\right]\Psi_r-\frac{t}{2}\sum_{\left<r,r'\right>}\Psi^{\dagger}_r[\tau_z\, e^{-i(\phi_r/2)\sigma_{z}}\sigma_0e^{i(\phi_{r'}/2)\sigma_{z}}]\Psi_{r'}.
\end{align}
This transformation aligns the spin at each site with the $x$-axis of the local spin basis~\cite{zhu_bogoliubov-gennes_2016, realistic_2D_Kitaev_model,PhysRevB.85.144505}.
To proceed, we focus on the gauge transformation that acts in the spin subspace of the kinetic term
\begin{align}
    \sum_{\left<r,r'\right>} e^{-i(\phi_r/2)\sigma_{z}}\sigma_0e^{i(\phi_{r'}/2)\sigma_{z}} &= \sum_{\left<r,r'\right>}\sigma_0\,\text{exp}\left[-\frac{i\sigma_z}{2}\left(\phi_r - \phi_{r'}\right)\right] = \sum_{\left<r,r'\right>}\sigma_0\,\text{exp}\left[\frac{i\sigma_z}{2}\left(2\pi q_x[r'_x-r_x] + 2\pi q_y[r'_y-r_y]\right)\right] \nonumber \\[0pt] 
    & = \sum_{\left<r,r'\right>}\sigma_0\,\text{exp}\left[i\sigma_z\pi\left( q_x\delta_{r_{x}+1,r'_x} + q_y\delta_{r_{y}+1,r'_y}\right)\right] = \sum_{\left<r,r'\right>} \sigma_0\,\text{exp}\left[i\sigma_z\pi \mathbf{q}\cdot\mathbf{v}_{r,r'}\right],
\end{align}
where $\mathbf{q} = (q_x,q_y)$ is the spiral pitch between adjacent spins and $\mathbf{v}_{r,r'} = (\delta_{r_{x}+1,r'_x}, \delta_{r_{y}+1,r'_y})$ is the real-space vector between the nearest neighbors $\langle r, r'\rangle$. The modified kinetic term is thus given by
\begin{align}
    H_{kin} &= \sum_{\left<r,r'\right>}\Psi^{\dagger}_r\left[-\frac{t}{2}\,\tau_z\,\exp(i\sigma_z\pi\,\mathbf{q}\cdot\mathbf{v}_{r,r'})\right]\Psi_{r'} \label{app_eq:MFB_stability_eff_SOC} \\ 
    &= \sum_{\left<r,r'\right>}\Psi^{\dagger}_r\left[-\frac{t}{2}\,\tau_z\sigma_0\cos(\pi\mathbf{q}\cdot\mathbf{v}_{r,r'}\,)-\frac{t}{2}\,\tau_z\,i\sigma_z\sin(\pi\mathbf{q}\cdot\mathbf{v}_{r,r'}\,)\right]\Psi_{r'},
\end{align}
where the sinusoidal component gives an effective spin-orbit interaction and the cosine term modifies the hopping between lattice sites. By substituting Eq.~\eqref{app_eq:MFB_stability_eff_SOC} in Eq.~\eqref{app_eq:MFB_stability_UdaggerHU}, we obtain the Hamiltonian in Eq.~\eqref{eq:gauge_transformed_H} in the main text.

\section{Derivation of winding number}
\label{app:winding_number}
For the $k$-space Hamiltonian $H(\mathbf{k})$ in Eq.~\eqref{eq:Fourier_transform_H}, the $k_y$-dependent winding number is given by an integral over $k_x$ as~\cite{PhysRevB.83.224511, PhysRevB.85.035110, PhysRevLett.109.150408}  
\begin{align}
    W(k_y) = -\frac{1}{4\pi i}\int_0^{2\pi} dk_x~\mathrm{tr}\left[SH^{-1}(\mathbf{k})\partial_{k_x}H(\mathbf{k})\right].
\end{align}
By choosing a basis in which the chiral symmetry operator $S$ is diagonal in the particle-hole subspace, the anticommutation $\{S,H\}=0$ implies that the Hamiltonian must take a block off-diagonal form
\begin{align}
    S = \begin{pmatrix}\sigma_0 & 0 \\ 0 & -\sigma_0\end{pmatrix}, \hspace{0.5cm} H(\mathbf{k}) = \begin{pmatrix}0 & A(\mathbf{k}) \\ A^{\dagger}(\mathbf{k}) & 0\end{pmatrix}. \nonumber
\end{align}
Consequently, $W(k_y)$ can be expressed in terms of $A(\mathbf{k})$ as
\begin{align}\label{eq:winding_number_det_A_form}
     W(k_y)=\frac{1}{2\pi}\mathrm{Im}\left[\int_0^{2\pi}dk_x~ \partial_{k_x}\ln{\det A(\mathbf{k})}\right].
\end{align}
To determine $A(\mathbf{k})$, we thus transform the $k$-space Hamiltonian in Eq.~\eqref{eq:Fourier_transform_H} into a block off-diagonal form in the particle-hole subspace using the unitary operator 
\begin{align}
    U=\frac{1}{2}\left(\tau_0\sigma_a+\tau_x\sigma^{*}_a+i\tau_y\sigma^{*}_b+\tau_z\sigma_b\right), \text{ where }\sigma_a = e^{i(\pi/4)}\begin{bmatrix}1 & 1 \\0 & 0\end{bmatrix},~\sigma_b = e^{i(\pi/4)}\begin{bmatrix}0 & 0 \\-1 & 1\end{bmatrix}.\nonumber
\end{align}
The transformed Hamiltonian is then
\begin{align}
    U^{\dagger}H(\mathbf{k})U = \begin{bmatrix}
    0 & \Delta_{\rm bulk}\,\sigma_x-i\left[\left(\mathcal{E}(\mathbf{k})-\mu\right)\,\sigma_0 +B\,\sigma_z + f(\mathbf{k})\,\sigma_x \right]  \\
    \Delta_{\rm bulk}\,\sigma_x+i\left[\left(\mathcal{E}(\mathbf{k})-\mu\right)\,\sigma_0 +B\,\sigma_z + f(\mathbf{k})\,\sigma_x \right] & 0  
    \end{bmatrix},
\end{align}
 which gives $A(\mathbf{k}) = \Delta_{\rm bulk}\,\sigma_x-i\left[\left(\mathcal{E}(\mathbf{k})-\mu\right)\,\sigma_0 +B\,\sigma_z + f(\mathbf{k})\,\sigma_x \right]$. 
 
 Next, by following the procedure outlined in Ref.~\cite{PhysRevB.83.224511}, we evaluate the integral \eqref{eq:winding_number_det_A_form} with $\det A(\mathbf{k}) = B^2-\left(\mathcal{E}(\mathbf{k})-\mu\right)^2-\left(\Delta_{\rm bulk}-if(\mathbf{k})\right)^2$. For brevity, we set $T(\mathbf{k}) = \mathcal{E}(\mathbf{k})-\mu$. This gives
 \begin{align}
    W(k_y)&=\frac{1}{2\pi}\mathrm{Im}\left[\int_0^{2\pi}dk_x~ \partial_{k_x}\ln{\left[B^2-T^2(\mathbf{k})-\Delta_{\rm bulk}^2+f^2(\mathbf{k})-2i\Delta_{\rm bulk} f(\mathbf{k})\right]}\right],\\[8pt]
    &=\frac{1}{2\pi}\int_0^{2\pi}dk_x~\frac{2\Delta_{\rm bulk} f(\mathbf{k})\partial_{k_x}\left[B^2-T^2(\mathbf{k})-\Delta_{\rm bulk}^2+f^2(\mathbf{k})\right] - \left[B^2-T^2(\mathbf{k})-\Delta_{\rm bulk}^2+f^2(\mathbf{k})\right]\partial_{k_x}\left[2\Delta_{\rm bulk} f(\mathbf{k})\right]}{\left[B^2-T^2(\mathbf{k})-\Delta_{\rm bulk}^2+f^2(\mathbf{k})\right]^2 + \left[2\Delta_{\rm bulk} f(\mathbf{k})\right]^2}.\label{app_eq:expanded_W_k_y}
 \end{align}

Since the winding number is invariant under continuous transformations, we may rescale either term in the denominator of the integrand to determine the leading order contribution to Eq.~\eqref{app_eq:expanded_W_k_y}. For simplicity, we rescale $\Delta_{\rm bulk}$ by taking $\Delta_{\rm bulk} \rightarrow \epsilon\Delta_{\rm bulk}$, where $\epsilon$ is a small positive constant. This gives   
\begin{align}\label{app_eq:rescaled_delta_W_k_y}
    &W(k_y)=\nonumber\\
    &\hspace{0.5cm}\frac{1}{2\pi}\int_0^{2\pi}dk_x~\frac{2\epsilon\Delta_{\rm bulk} f(\mathbf{k})\partial_{k_x}\left[B^2-T^2(\mathbf{k})-(\epsilon\Delta_{\rm bulk})^2+f^2(\mathbf{k})\right] - \left[B^2-T^2(\mathbf{k})-(\epsilon\Delta_{\rm bulk})^2+f^2(\mathbf{k})\right]\partial_{k_x}\left[2\epsilon\Delta_{\rm bulk} f(\mathbf{k})\right]}{\left[B^2-T^2(\mathbf{k})-(\epsilon\Delta_{\rm bulk})^2+f^2(\mathbf{k})\right]^2+\left[2\epsilon\Delta_{\rm bulk} f(\mathbf{k})\right]^2}.
\end{align}
Here the integrand sharply peaks near the zeros $k_x^0$ of the denominator $B^2-T^2(\mathbf{k})+f^2(\mathbf{k})$. Hence, in the limit of small $\epsilon\Delta_{\rm bulk}$, the dominant contribution to the integral comes from the neighborhood $\delta$ of these points, and the winding number is given by the sum of the contributions from all such zeros. By expanding the functions in the integral Eq.~\eqref{app_eq:rescaled_delta_W_k_y} near $k_x^0$, satisfying the condition $B^2-T^2(k_x^0,k_y)+f^2(k_x^0,k_y) = 0$, we arrive at the lowest order non-vanishing terms
\begin{align}
    &B^2-T^2(\mathbf{k})-(\epsilon\Delta_{\rm bulk})^2+f^2(\mathbf{k}) \approx \partial_{k_x}\left[B^2-T^2(k_x,k_y) - (\epsilon\Delta_{\rm bulk})^2 +f^2(k_x,k_y)\right]\Big|_{k_x = k_x^0}(k_x - k_x^0) + \dots\,, \\
    &f(\mathbf{k}) = f(k_x^0,k_y) + \dots\,.
\end{align}
As the term $\partial_{k_x}\left[2\epsilon\Delta_{\rm bulk} f(k_x^0,k_y)\right] =0$, the winding number can be estimated as
\begin{align}
    W(k_y) = \sum_{k_x^0}\frac{1}{2\pi}\int_{k_x^0-\delta}^{k_x^0+\delta}dk_x~\frac{2\epsilon\Delta_{\rm bulk} f(k_x^0,k_y)\partial_{k_x}\left[B^2-T^2(k_x^0,k_y)-(\epsilon\Delta_{\rm bulk})^2+f^2(k_x^0,k_y)\right] }{\left(\partial_{k_x}\left[B^2-T^2(k_x^0,k_y)-(\epsilon\Delta_{\rm bulk})^2+f^2(k_x^0,k_y)\right]\right)^2 (k_x-k^0_x)^2 + \left[2\epsilon\Delta_{\rm bulk} f(k_x^0,k_y)\right]^2}\,.
\end{align}
Now, setting the constant $\eta_{k^0_x} = \partial_{k_x}\left[B^2-T^2(k_x,k_y)-(\epsilon\Delta_{\rm bulk})^2+f^2(k_x,k_y)\right]\Big|_{k_x = k_x^0}$ for brevity and upon a change of variable $z=(k_x-k^0_x)$, we obtain
\begin{align}
    W(k_y) = \sum_{k_x^0}\frac{2\epsilon\Delta_{\rm bulk} f(k_x^0,k_y)\eta_{k^0_x}}{2\pi}\int_{-\delta}^{\delta}dz \frac{1}{\eta^2_{k^0_x}z^2+\left[2\epsilon\Delta_{\rm bulk} f(k_x^0,k_y)\right]^2} = \sum_{k_x^0}\frac{1}{\pi}\tan^{-1}\left(\frac{\eta_{k^0_x}\delta}{2\epsilon\Delta_{\rm bulk} f(k_x^0,k_y)}\right).
\end{align}
As the numerator is dominated by the leading-order term, $\eta_{k^0_x}\delta \gg 2\epsilon\Delta_{\rm bulk} f(k_x^0,k_y)$, the argument of the arctangent approaches infinity. By setting the positive constant $\epsilon$ to unity, the winding number can be written as
\begin{align}\label{app_eq:simplified_W_k_y_eq_1}
    W(k_y) &= \frac{1}{2} \sum_{k_x^0} \mathrm{sgn}\left[\partial_{k_x}\left[B^2-T^2(k_x,k_y)-\Delta_{\rm bulk}^2+f^2(k_x,k_y)\right]\Big|_{k_x = k_x^0}\right]\mathrm{sgn}\left[\Delta_{\rm bulk} f(k_x^0,k_y)\right],
\end{align}
where $k^0_x$ are the points that satisfy the condition $B^2-T^2(k_x^0,k_y)+f^2(k_x^0,k_y) = 0$, at fixed values of $k_y$.\\[0.08cm]
An equivalent expression for the winding number can be obtained by scaling the term $[B^2-T^2(\mathbf{k})-\Delta_{\rm bulk}^2+f^2(\mathbf{k})]$ with $\epsilon$ instead of $\Delta_{\rm bulk}$ in the denominator of the integrand in Eq.~\eqref{app_eq:expanded_W_k_y}. Following the same procedure, this effectively interchanges the roles of $\Delta_{\rm bulk} f(k_x^0,k_y)$ and $B^2-T^2(k_x,k_y)-\Delta_{\rm bulk}^2+f^2(k_x,k_y)$ in Eq.~\eqref{app_eq:simplified_W_k_y_eq_1} giving
\begin{align}\label{app_eq:simplified_W_k_y_eq_2}
    W(k_y) = -\frac{1}{2} \sum_{k_x^0} \mathrm{sgn}\left[\partial_{k_x}\left[\Delta_{\rm bulk} f(k_x,k_y)\right]\Big|_{k_x = k_x^0}\right]\mathrm{sgn}\left[B^2-T^2(k_x^0,k_y)-\Delta_{\rm bulk}^2+f^2(k_x^0,k_y)\right],
\end{align}
where $k^0_x$ are now the points that satisfy the condition $f(k_x^0,k_y) = 0$, at fixed values of $k_y$. The expression in Eq.~\eqref{app_eq:simplified_W_k_y_eq_2} provides a simple and straightforward evaluation of the winding numbers. By further re-writing, we finally arrive at 
\begin{align}
    W(k_y) = -\frac{1}{2}\sum_{f(\mathbf{k})=0} \mathrm{sgn}\left[\Delta_{\rm bulk}\,\partial_{k_x}f(\mathbf{k})\right]\mathrm{sgn}\left[B^2-\left[\mathcal{E}(\mathbf{k})-\mu\right]^2-\Delta_{\rm bulk}^2+f^2(\mathbf{k})\right],
\end{align}
which is given as Eq.~\eqref{eq:W_ky_formula} in the main text.

\begin{figure*}[t!]
    \centering
    \includegraphics[width=1\linewidth]{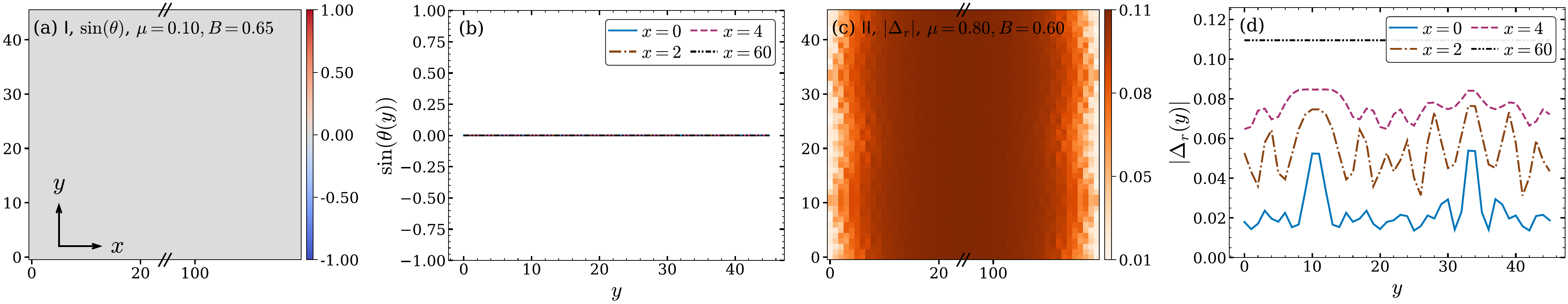}
    \caption{Pair density wave state (a,b) and phase crystal state (c,d), corresponding to points I at $(\mu, B) = (0.10, 0.65)$ and II at $(\mu, B) = (0.80, 0.60)$, in Fig.~\ref{fig:zero_temp_phase_diagram}, respectively. (a,c) Color density maps of the superconducting order parameter phase $\sin\theta$ at I and amplitude $|\Delta_r|$ at II, respectively. (b,d) Corresponding line plots of the phase modulations and the amplitude modulations as a function of position $y$ along the edge for several fixed values of $x$ with $x=0$ being the edge and $x=60$ the bulk.}     
    \label{app_fig:representative_points_additional_data}
\end{figure*}

\begin{figure*}[t!]
    \centering
    \includegraphics[width=0.65\linewidth]{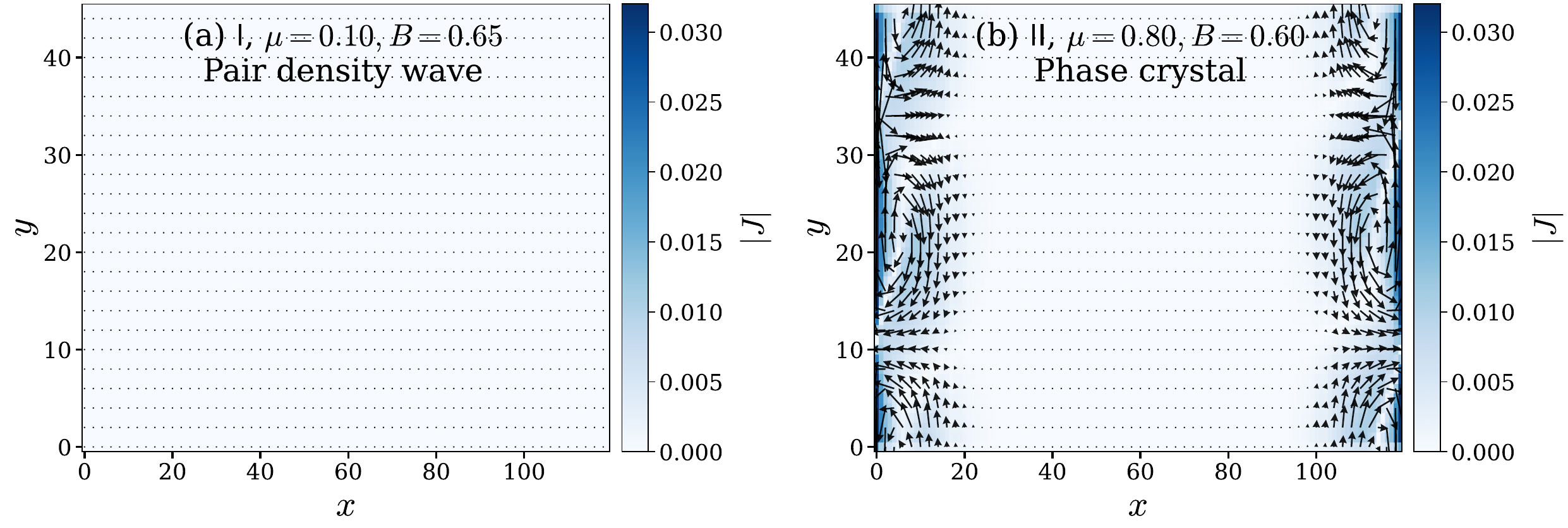}
    \caption{(a,b) Color density maps of the magnitude of current vector field $|J|$ and the corresponding vector plots for pair density wave state at I, $(\mu, B) = (0.10, 0.65)$ and for phase crystal state at II, $(\mu, B) = (0.80, 0.60)$, in Fig.~\ref{fig:zero_temp_phase_diagram}, respectively.}     
    \label{app_fig:representative_points_current_plots}
\end{figure*}

\begin{figure*}[t!]
    \centering
    \includegraphics[width=1\linewidth]{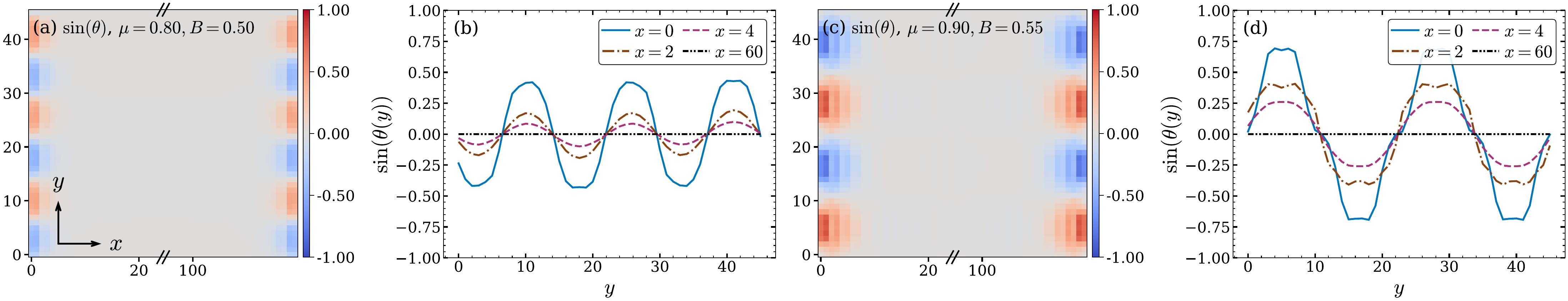}
    \caption{Additional data in the phase crystal state at zero temperature. (a,c) Color density maps of the superconducting order parameter phase $\sin\theta$ at $(\mu, B) = (0.80, 0.50)$ and $(0.90, 0.55)$, in Fig.~\ref{fig:zero_temp_phase_diagram}, respectively. (b,d) Corresponding line plots as a function of position $y$ along the edge for several fixed values of $x$.}     
    \label{app_fig:phase_crystal_additional_data}
\end{figure*}

\section{Additional data for pair density wave and phase crystal}
\label{app:additional_bwo_pc_data}
Figures~\ref{app_fig:representative_points_additional_data}\textcolor{blue}{(a,b)} show the vanishing phase modulations of the pair density wave at point I in Fig.~\ref{fig:zero_temp_phase_diagram}, corresponding to the parameters $\mu=0.10, B=0.65$, as discussed in Sec.~\ref{sec:Bond_wave_order}. Similarly, Figs.~\ref{app_fig:representative_points_additional_data}\textcolor{blue}{(c,d)} depict the weak amplitude modulations driven by the phase crystal at point II in Fig.~\ref{fig:zero_temp_phase_diagram} for the parameters $\mu=0.80, B=0.60$, as discussed in Sec.~\ref{sec:Phase_crystal}. Figure~\ref{app_fig:representative_points_current_plots}\textcolor{blue}{(a)} shows the absence of currents in the pair density wave state at point I due to lack of phase modulations. In contrast, Fig.~\ref{app_fig:representative_points_current_plots}\textcolor{blue}{(b)} shows the spontaneous staggered loop currents in the phase crystal state at point II that break both translation and time-reversal symmetries, as discussed in Sec.~\ref{sec:Phase_crystal}. Furthermore, to provide an overview of different phase crystal configurations obtained as ground states at zero temperature, we present additional representative data in Fig.~\ref{app_fig:phase_crystal_additional_data} at two other points in the phase diagram, as discussed in Sec.~\ref{sec:Phase_crystal}.

\section{Additional data for lattice sizes with fixed $N_x$ and varying $N_y$} 
\label{app:additional_lattice_size_data}
For a few representative parameters in the zero-temperature phase diagram (Fig.~\ref{fig:zero_temp_phase_diagram}), we perform additional calculations, fixing $N_x = 40$ and varying $N_y$ from $40$ to $160$. This is to verify that the results are not an artifact of a finite $N_y$.
For this data, we compute the dominant frequency component $q_{\mathrm{max}}$, corresponding to the maximum intensity in the Fourier power spectrum, for both the amplitude and the phase along the edge at several fixed values of $x$ for the converged solution with the lowest free energy. From this we extract the spatial wavelengths $\lambda = 1/q_{\mathrm{max}}$ of the amplitude ($\lambda_{|\Delta_r|}$) and phase ($\lambda_{\theta_r}$) modulations along the edge ($y$-direction).

\begin{figure*}[t!]
    \centering
    \includegraphics[width=1\linewidth]{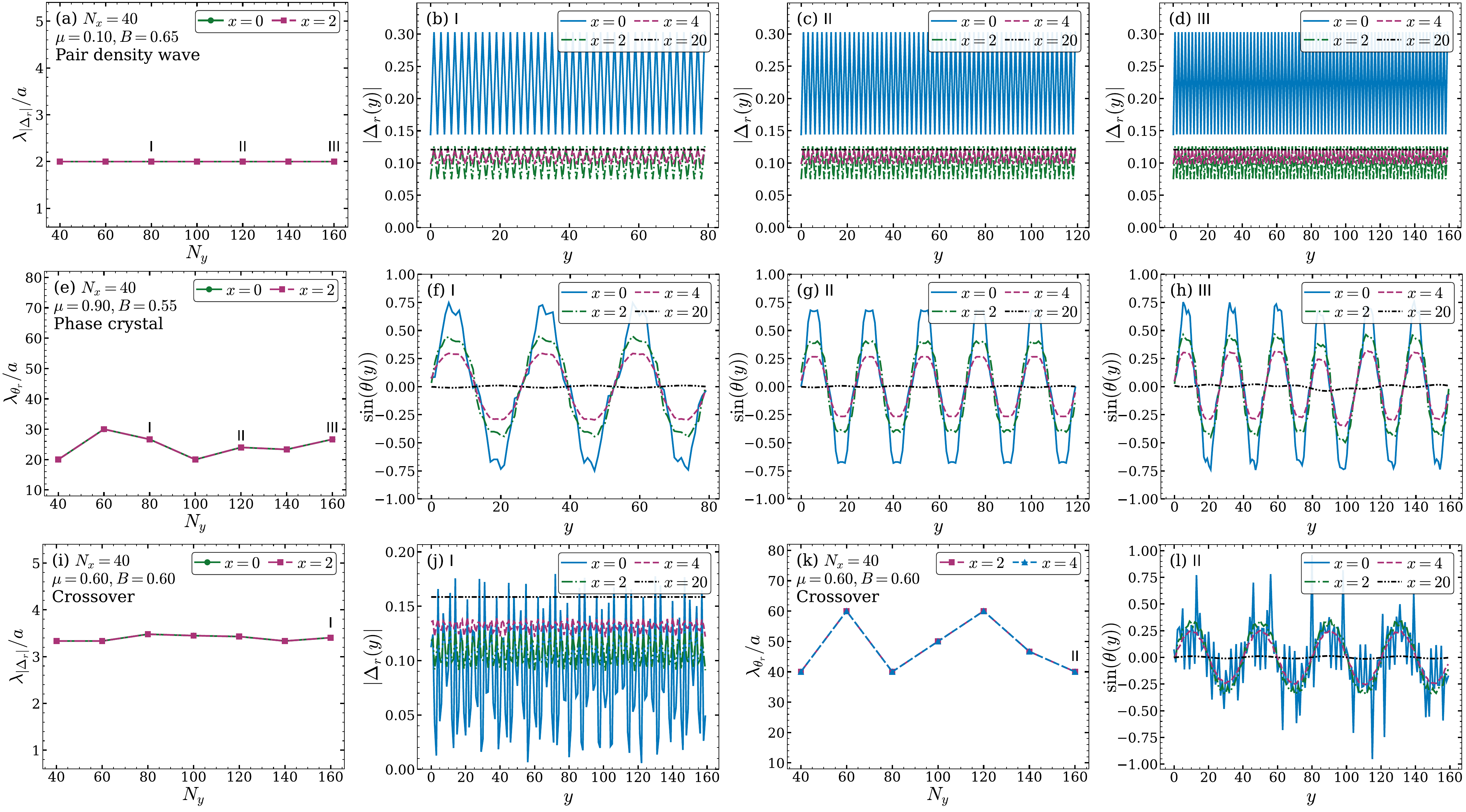}
    \caption{Pair density wave state [top row, (a-d)], phase crystal state [middle row, (e-h)], and a crossover state [bottom row, (i-l)] at zero temperature, corresponding to parameters $(\mu, B) = (0.10, 0.65)$, $(0.90, 0.55)$, and $(0.60, 0.60)$, respectively. Spatial wavelength of the amplitude modulation (a, i) and phase modulations (e, k) along the edge ($y$-direction) as a function of $N_y$ for fixed $N_x = 40$. (b-d), (f-h), and (j,l): Corresponding line plots of amplitude and phase modulations as a function of position $y$ along the edge for several fixed values of $x$ with $x=0$ being the edge and $x=20$ representing the bulk.}       
    \label{app_fig:additional_lattice_size_data}
\end{figure*}

The pair density wave, here extracted at $(\mu, B) = (0.10, 0.65)$, has a modulation period $\lambda_{|\Delta_r|}=2a$, i.e.,~a bond-wave order, for all different values of $N_y$, as shown in Fig.~\ref{app_fig:additional_lattice_size_data}\textcolor{blue}{(a)}. Line plots at specific $x$ values, ranging from the edge $x=0$ to the interior $x=20$, of the amplitude modulation at points I-III in Fig.~\ref{app_fig:additional_lattice_size_data}\textcolor{blue}{(a)} are shown in Figs.~\ref{app_fig:additional_lattice_size_data}\textcolor{blue}{(b-d)}. Similarly, Figs.~\ref{app_fig:additional_lattice_size_data}\textcolor{blue}{(e-h)} present the phase modulation period and line plots of the phase crystal at $(\mu, B) = (0.90, 0.55)$. The phase crystal acquires a periodicity of $\lambda_{\theta_r}\sim 20a-30a$, showing only small finite-size corrections for varying $N_y$. Finally, as shown in Figs.~\ref{app_fig:additional_lattice_size_data}\textcolor{blue}{(i,k)}, the crossover state
at $(\mu, B) = (0.60, 0.60)$ inherits features from both phases, exhibiting $\lambda_{|\Delta_r|} \approx 3.4a$, and $\lambda_{\theta_r} \sim 40a - 60a$ across $N_y$. Figures~\ref{app_fig:additional_lattice_size_data}\textcolor{blue}{(j,l)} further show the corresponding line plots for amplitude and phase modulations at $N_y = 140$. 
This establishes that a finite $N_y$ has a minimal effect on the spatial modulations.
We note that the small finite-size corrections we find in the phase modulations stem from incommensurability effects, as the phase crystal and crossover states need to accommodate pairs of counterflowing current loops along the edge.

\clearpage

\twocolumngrid

\begin{thebibliography}{98}
\bibitem [{cro()}]{crossover_footnote}%
  \BibitemOpen
  \href@noop {} {}\bibinfo {note} {Immediately after the system enters the
  topologically non-trivial phase, specifically at $B=0.45$ and $0.50$, we need
  to halve the bounds on the mean RMS phase deviation $\delta\theta$ for
  consistent behavior. This reduction primarily accounts for the lower number
  of Majorana states in this parameter range compared to the rest of the phase
  diagram, leading to not needing as large phase variations to shift states
  away from zero energy.}\BibitemShut {Stop}%
\end{thebibliography}
%

\end{document}